\documentclass[10pt,preprint]{aastex}

\newcommand\kms           {km~s$^{-1}$}
\newcommand\lsr           {\ensuremath{_{\mathrm {LSR}}}}
\newcommand\phmin         {\phm{$-$}}
\newcommand{\rarr}        {\ensuremath{\rightarrow}}
\newcommand{\tnm}         {\tablenotemark}
\newcommand{\tnt}         {\tablenotetext}
\newcommand{\hii}         {\ion{H}{2}}
\newcommand{\ttt}          {\ensuremath{3^+ \rarr 3^-}}
\newcommand{\ttf}          {\ensuremath{3^+ \rarr 4^-}}
\newcommand{\tft}          {\ensuremath{4^+ \rarr 3^-}}
\newcommand{\tff}          {\ensuremath{4^+ \rarr 4^-}}
\newcommand{\threehalves} {$^2\Pi_{3/2},\, J = 3/2$}
\newcommand{\fivehalves}  {$^2\Pi_{3/2},\, J = 5/2$}
\newcommand{\sevenhalves} {$^2\Pi_{3/2},\, J = 7/2$}
\shorttitle{Magnetic Field Clumping From OH Absorption}
\shortauthors{Fish et al.}
\begin{document}

\title{Magnetic Field Clumping in Massive Star-Forming Regions as
  Determined from Excited-State OH Absorption and Maser Emission}
\author{Vincent~L.~Fish\altaffilmark{1}}
\affil{National Radio Astronomy Observatory}
\affil{P.O.~Box O, 1003 Lopezville Road, Socorro, NM 87801}
\email{vfish@nrao.edu}
\author{Mark~J.~Reid}
\affil{Harvard--Smithsonian Center for Astrophysics}
\affil{60 Garden Street, Cambridge, MA 02138}
\email{reid@cfa.harvard.edu}
\and
\author{Karl~M.~Menten}
\affil{Max-Planck-Institut f\"{u}r Radioastronomie}
\affil{Auf dem H\"{u}gel 69, Bonn, D-53121 Germany}
\email{kmenten@mpifr-bonn.mpg.de}

\altaffiltext{1}{Jansky Fellow}

\begin{abstract}
We have observed six high-mass star-forming regions in the
\sevenhalves\ lines of OH using the GBT in order to investigate
whether the magnetic field, and hence the density, measured in
absorption differs from that implied by maser Zeeman splitting.  We
detect absorption in both the 13441 and 13434~MHz main lines in all
six sources.  Zeeman splitting in the $F = \ttt$ absorption line in
W3(OH) implies a line-of-sight magnetic field strength of $3.0 \pm
0.3$~mG.  This is significantly less than full magnetic field
strengths detected from OH maser Zeeman splitting, suggesting that OH
maser regions may be denser than the non-masing OH material by a
factor of several.  Zeeman splitting is not detected in other sources,
but we are able to place upper limits on $B_\parallel$ of 1.2~mG in
G10.624$-$0.385 and 2.9~mG in K3$-$50.  These results are consistent
with a density enhancement of the masers, but other explanations for
the lower magnetic field in absorption compared to maser emission are
possible for these two sources.  Absorption in one or both of the
13442 and 13433~MHz satellite lines is also seen in four sources.
This is the very first detection of the \sevenhalves\ satellite lines.
Ratios of satellite-line to main-line absorption suggest enhancement
of the satellite lines from local thermodynamic equilibrium values.
Masers are seen in the $F = \tff$ and $\ttt$ transitions of W3(OH) and
the $\tff$ transition of ON~1.  A previously undetected $\tff$ maser
is seen near $-44.85$~\kms\ in W3(OH).
\end{abstract}


\section{Introduction}

Hydroxyl (OH) masers are found in high-mass star-forming regions.
They trace the local velocity and magnetic field and therefore provide
clues to understanding the physical conditions of material surrounding
newly-formed high-mass stars.  But an unanswered question is whether
the physical conditions in masing regions are representative of the
surrounding material.  In order for OH masing to occur, a large column
density of OH must have velocity coherence such that the velocity
gradient along the amplification path of the masing clump does not
exceed the masing linewidth.  In principle, sufficient column density
can be achieved in two ways.  If a masing clump is much denser than
the ambient gas, the total OH column density can be large even if the
physical extent of the clump is not.  Alternatively, masing may occur
along favored paths of velocity coherence even in a medium of
homogeneous density.

While it is uncertain which of these two scenarios is prevalent, two
pieces of evidence suggest that density enhancements may not be
necessary for OH masing to occur.  First, 13434 MHz OH
\emph{absorption} in W3(OH) shows Zeeman splitting indicating a
line-of-sight magnetic field strength of 3.1 mG \citep{gusten}.  VLBI
measurements of 1665 and 1667 MHz OH \emph{masers} in W3(OH) imply
similar magnetic field strengths \citep[e.g.,][]{bloemhof}.  The
magnetic field strength of collapsing (or collapsed) material scales
as the density $n^{\kappa}$, where $\kappa \approx 0.5$, a result
supported both by theoretical modelling \citep{mouschovias,fiedler93}
and observations of molecular clouds in various stages of collapse
\citep[e.g.,][]{crutcher91}.
Since the OH absorption and
maser magnetic fields are comparable, this suggests that the density
at masing sites is similar to that of the ambient cloud of OH.
Second, ammonia (NH$_3$) observations of W3(OH)
indicate that the density of material in the clumps of maser emission
is roughly the same (to within a factor of two) as the density of the
interclump material \citep{rmb}.  Since the velocity and extent of
NH$_3$ absorption is similar to that of the OH emission in W3(OH), it
is a reasonable assumption that NH$_3$ and OH exist in the same cloud
of material.
Finally, \citet{cesaroni}, when modeling multi-transition
OH observations of
W3(OH), find that maser emission in certain lines and absorption
in others can be explained for the same range
of densities, between $10^6$ and a few times $10^7$ cm$^{-3}$
(for a temperature of 150 K).

Is W3(OH) a special case, or is it representative of all interstellar
OH maser sources?  In particular, is any density enhancement required
for the onset of OH masing?  These are questions that motivated us to
observe a wider range of sources than \citet{gusten} with the high
sensitivity that the GBT can afford.

\section{Observations}

The observations were performed on 2004 April 11 and 12 using the
National Radio Astronomy Observatory's\footnote{The National Radio
Astronomy Observatory is a facility of the National Science Foundation
operated under cooperative agreement by Associated Universities, Inc.}
Robert C. Byrd Green Bank Telescope (GBT) in Green Bank, WV.  The GBT
has an effective diameter of 100 m.  Observations were taken in both
circular polarizations with the Gregorian focus Ku-band receiver.  The
GBT Spectrometer was configured in 9-level mode to provide 8192
uniform-weighted spectral channels in each of two IFs covering a
bandwidth of 12.5 MHz each, centered on 13441.4173 and 13434.6374 MHz,
Doppler-shifted to the LSR velocity of each source.  These correspond
to the frequencies of the $F = \tff$ and $F = \ttt$ main-line
\sevenhalves\ transitions, respectively \citep{destombes}.  Each
spectral channel had a resolution of 1.5 kHz, corresponding to a
resolution of 0.034 \kms\ in velocity space.  Thus, the 13433.982 MHz
($F = \tft$) and 13442.072 MHz ($F = \ttf$) satellite lines were also
in the observed frequency range.  At these frequencies, the FWHM
beamwidth of the GBT is 55\arcsec\ and the gain is 1.7 K/Jy.

Double beam switching was employed such that the source appeared in
each of the dual beams alternately for a period of two minutes.  The
beams were separated by 330\arcsec\ in azimuth.  Dynamic pointing and
focusing were used, and the pointing and focus were checked hourly as
well as each time the telescope was pointed at a source.

Variable rain, heavy at times, fell throughout the data collection
period.  During the brief interludes without precipitation, the
(non-zenith) system temperature approached being receiver-limited at
about 30~K.  At times, the system temperature was over 100~K, due to
heavy precipitation and low elevation angles.

Our sources were chosen according to two criteria.  First, they must
contain a strong background \hii\ region so that absorption might be
observed.  Second, since it was our intent to compare the magnetic
field strength in the OH gas seen in absorption with that seen in
maser emission, we sought sources in which the masers indicated a
uniform field distribution and eliminated any with a reversal of the
line-of-sight direction of the magnetic field as determined from
ground-state OH maser Zeeman splitting \citep{fishthesis}.
Only about 10 sources observable at the latitude of the GBT meet
these criteria.  We observed the six sources listed in Table
\ref{source-table}.

\begin{deluxetable}{lllll}
\tablewidth{0 pt}
\tablecaption{Observed Sources\label{source-table}}
\tablehead{ \colhead{}&\colhead{}&\colhead{}&
            \colhead{Obs.\ Time} &
            \colhead{$\sigma$\tnm{a}} \\
            \colhead{Source} &
            \colhead{RA(J2000)} &
            \colhead{Dec(J2000)} &
            \colhead{(min)} &
            \colhead{(K)}
          }
\startdata
W3(OH)          & 02 27 03.70 & $+$61 52 25.4 & 276 & 0.006 \\
G10.624$-$0.385 & 18 10 28.61 & $-$19 55 49.7 & 270 & 0.008 \\
G28.199$-$0.048 & 18 42 58.04 & $-$04 13 58.0&\phn10& 0.026 \\
W49             & 19 10 11.04 & $+$09 05 20.2&\phn58& 0.028 \\
K3$-$50         & 20 01 45.73 & $+$33 32 45.3 & 178 & 0.010 \\
ON 1            & 20 10 09.05 & $+$31 31 35.2&\phn20& 0.015 \\
\enddata
\tnt{a}{Single-channel RMS noise in Stokes I and V.}
\end{deluxetable}

\section{Results}

\begin{figure}
\begin{center}
\includegraphics[width=4.0in]{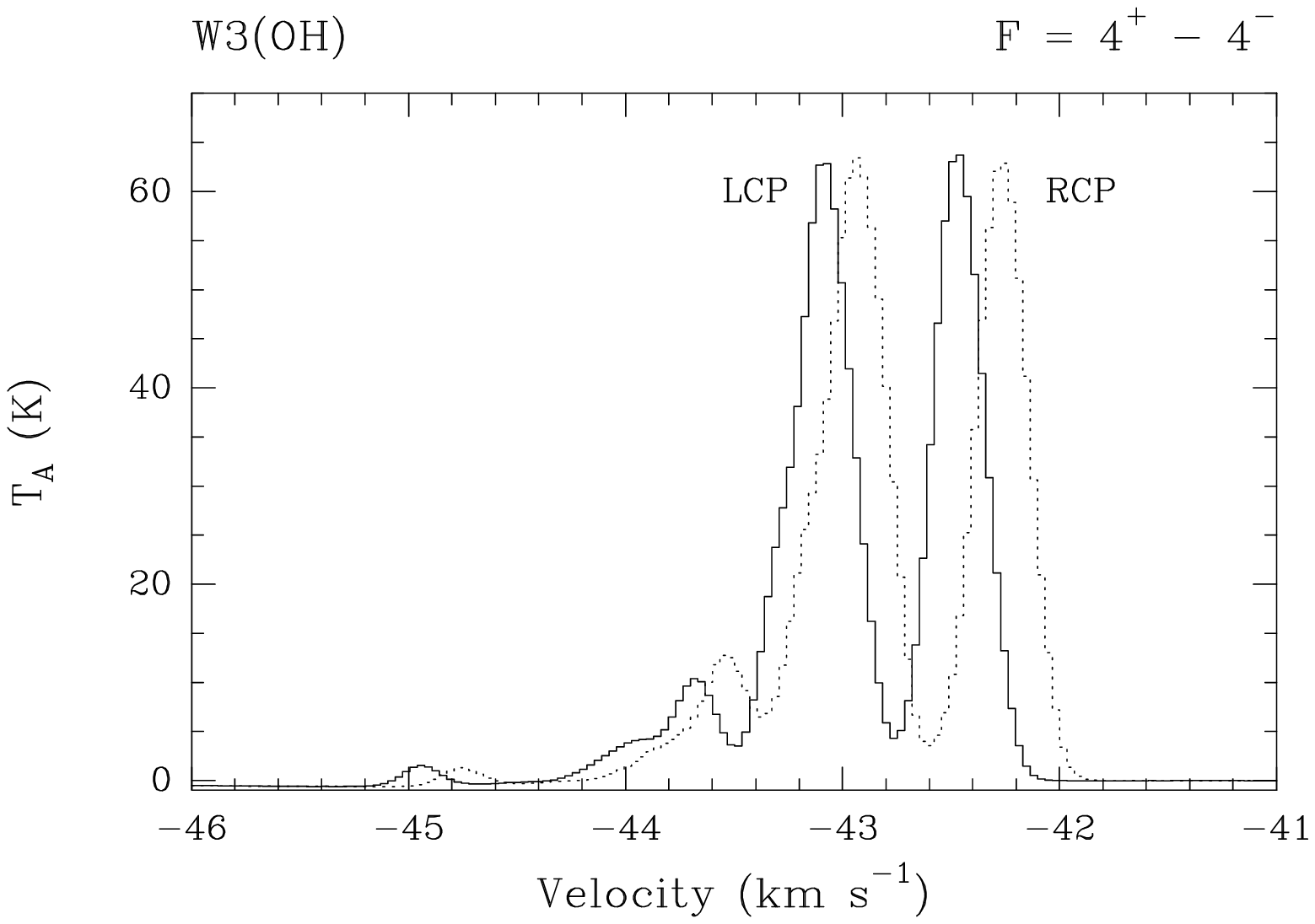}
\end{center}
\caption{Masers in the $F = \tff$ transition of W3(OH).  LCP emission
  is plotted as a solid line, and RCP emission is plotted as a dotted
  line.  Maser fit parameters are given in Table
  \ref{maser-table}.\label{w3ohlr1}}
\end{figure}

\begin{figure}
\begin{center}
\includegraphics[width=4.0in]{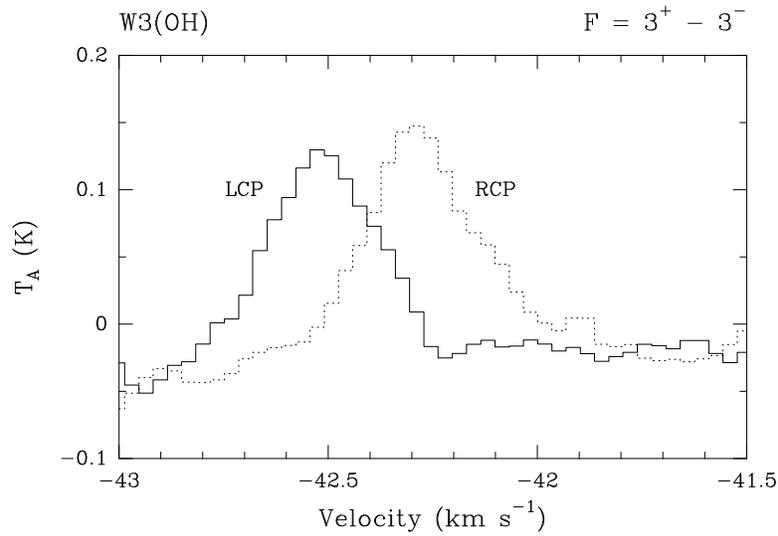}
\end{center}
\caption{Masers in the $F = \ttt$ transition of W3(OH).  See Figure
  \ref{w3ohlr1} caption for details.\label{w3ohlr2}}
\end{figure}

\begin{figure}
\begin{center}
\includegraphics[width=4.0in]{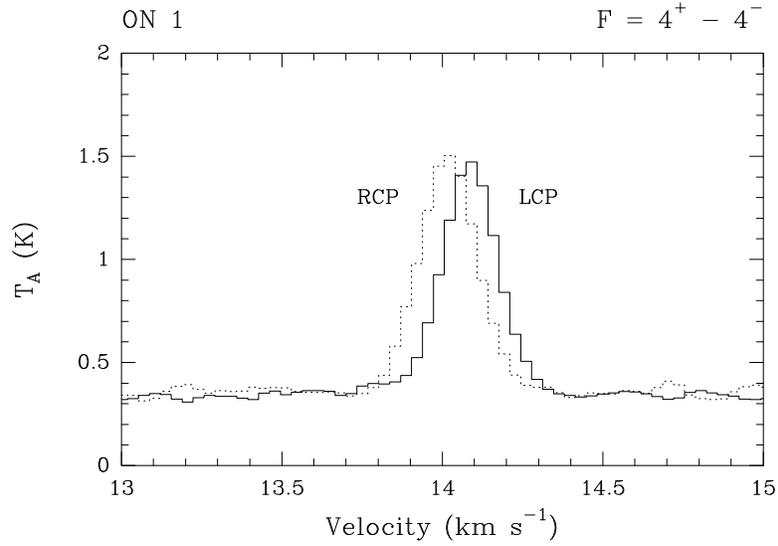}
\end{center}
\caption{Masers near $14$~\kms\ in the $F = \tff$ transition of ON~1.
  See Figure \ref{w3ohlr1} caption for details.\label{on1lr1}}
\end{figure}

\begin{figure}
\begin{center}
\includegraphics[width=4.0in]{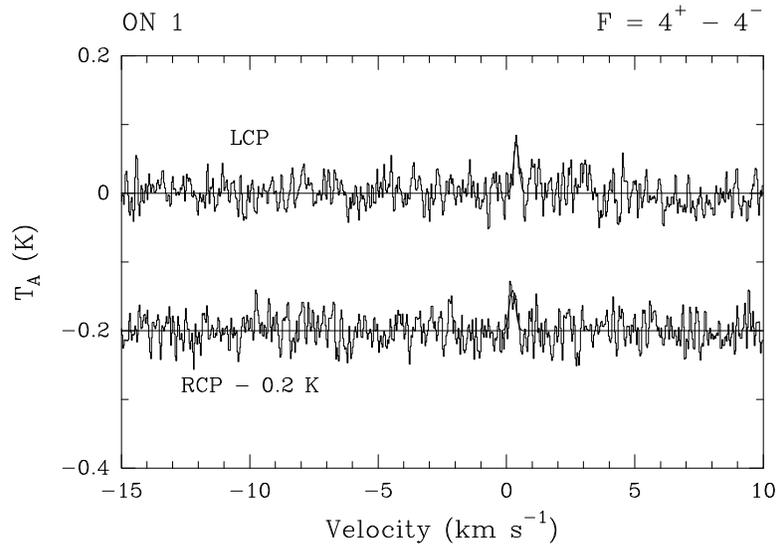}
\end{center}
\caption{Masers near $0$~\kms\ in the $F = \tff$ transition of ON~1.
  The LCP and RCP data are plotted as histograms, and the best
  Gaussian fits are plotted as curves.  The RCP data have been shifted
  by 0.2~K for clarity.\label{on1lr1a}}
\end{figure}

\begin{figure}
\begin{center}
\includegraphics[width=4.0in]{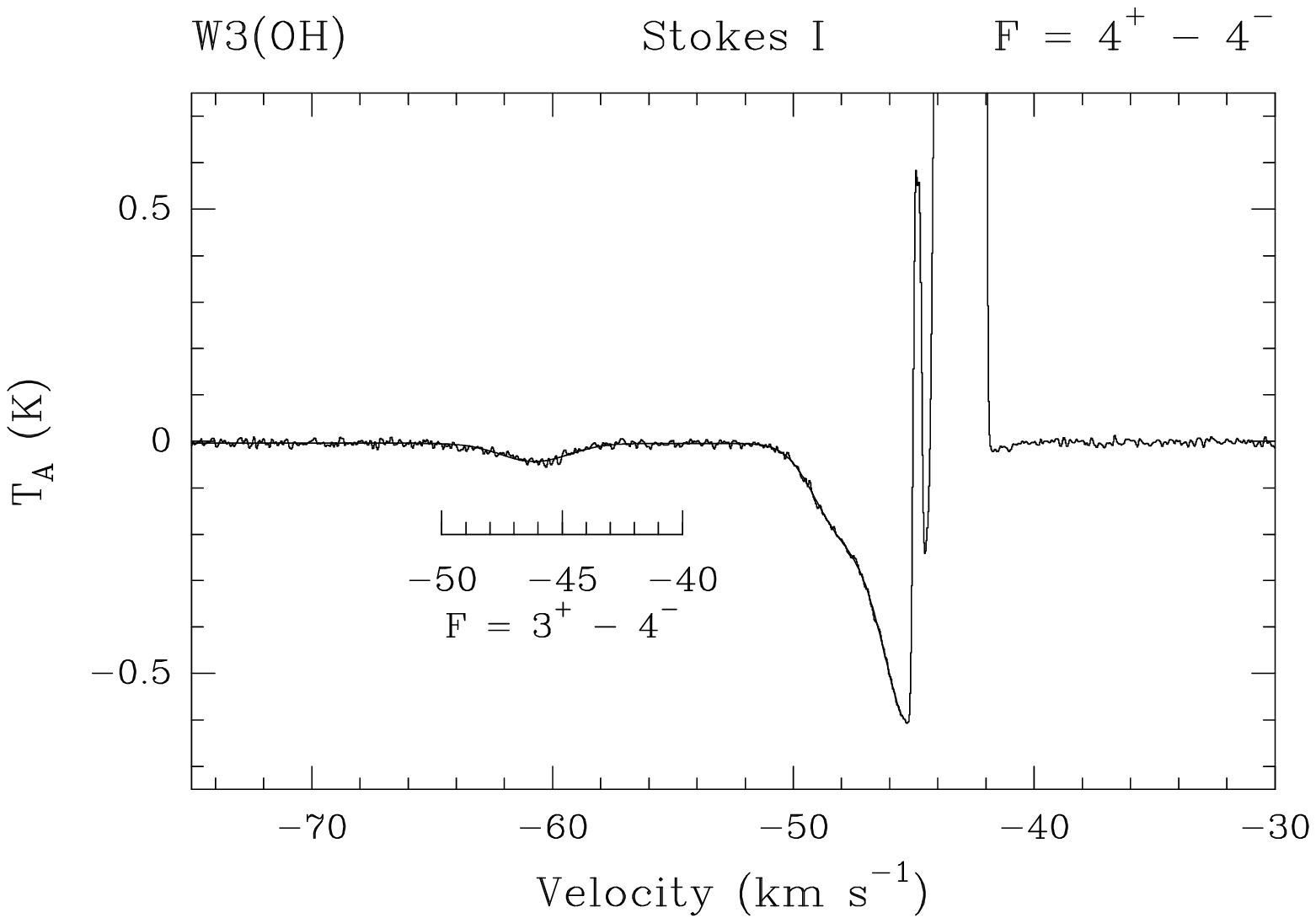}
\end{center}
\caption{Stokes I spectrum for the $F = \tff$ and $\ttf$ transtitions
  of W3(OH).  Data points are Hanning weighted and plotted as a
  histrogram.  The Gaussian fit parameters listed in Table
  \ref{absorption-table} are plotted as a curve.  The masers are
  plotted in Figure \ref{w3ohlr1} with a much larger scale to show the
  strong emission.  See \S \ref{absorption} for details regarding the
  fits to the main-line absorption features.  The $F = \ttf$ line is
  shown in greater detail in Figure \ref{w3ohi1a}.
  \label{w3ohi1}}
\end{figure}

\begin{figure}
\begin{center}
\includegraphics[width=4.0in]{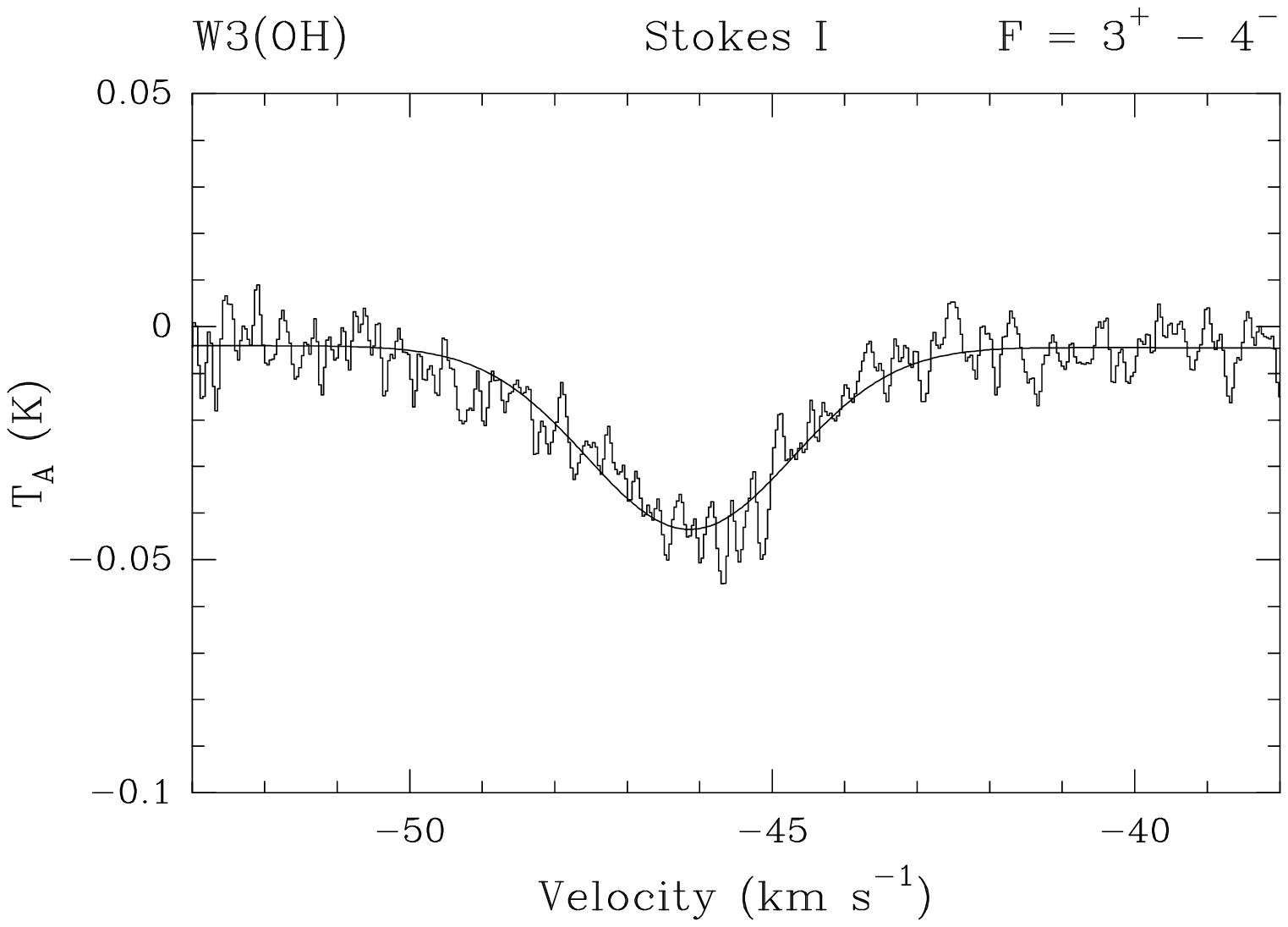}
\end{center}
\caption{Stokes I spectrum for the $F = \ttf$ transtition of W3(OH).
  See Figure \ref{w3ohi1} caption for more details.\label{w3ohi1a}}
\end{figure}

\begin{figure}
\begin{center}
\includegraphics[width=4.0in]{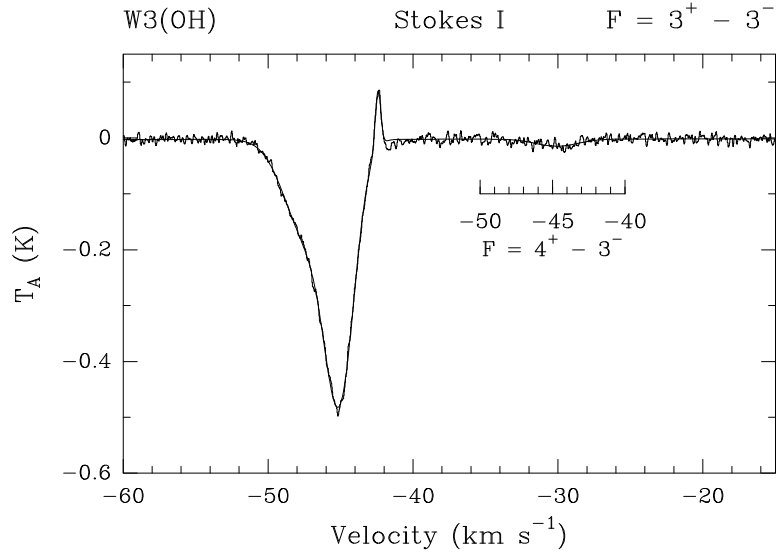}
\end{center}
\caption{Stokes I spectrum for the $F = \ttt$ and $\tft$ transitions
  of W3(OH).  See Figure \ref{w3ohi1} caption for more details.  The
  maser at $-42$~\kms\ is plotted in Figure
  \ref{w3ohlr2}.  The $F = \tft$ line is
  shown in greater detail in Figure \ref{w3ohi2a}.\label{w3ohi2}}
\end{figure}

\begin{figure}
\begin{center}
\includegraphics[width=4.0in]{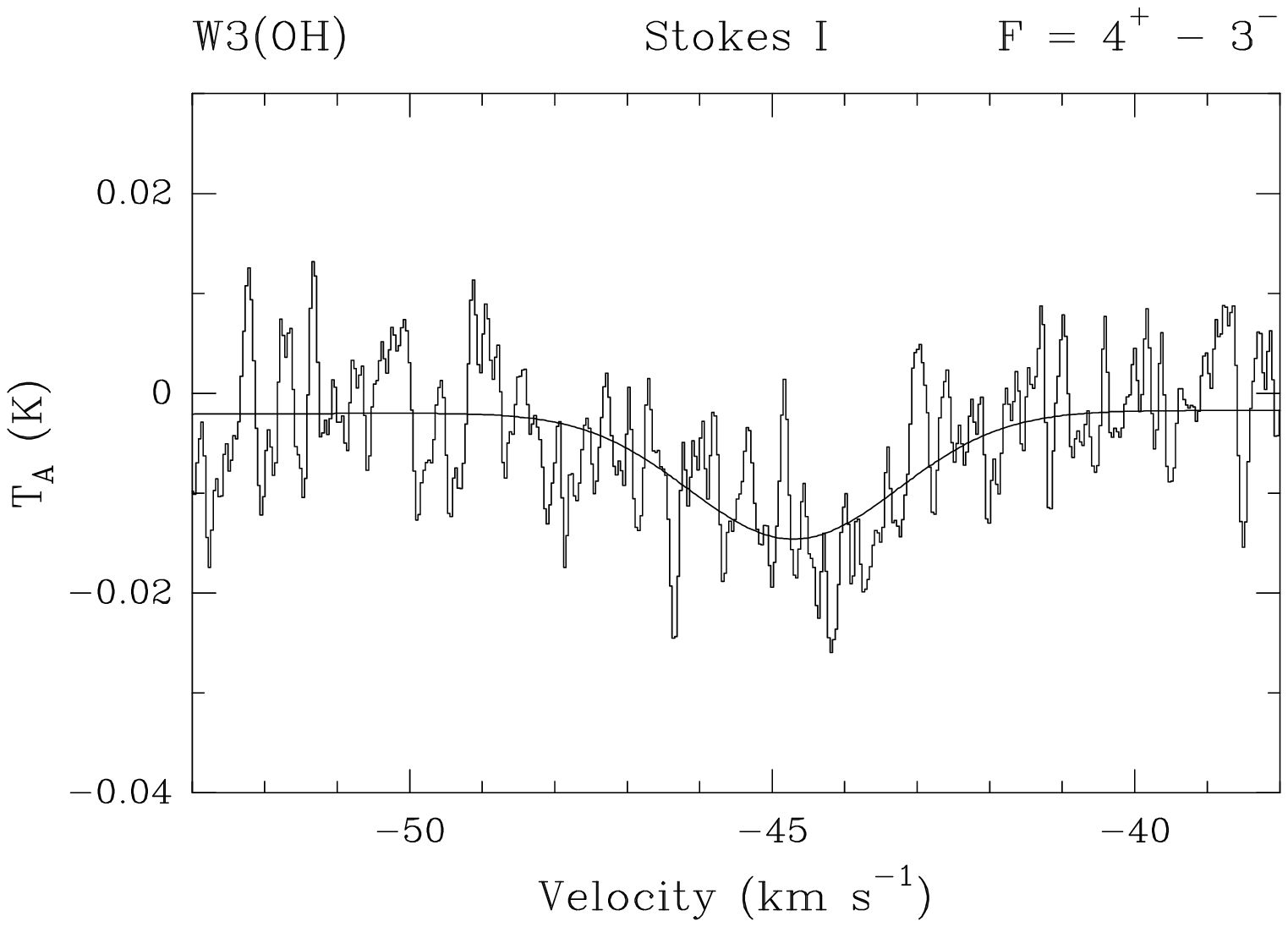}
\end{center}
\caption{Stokes I spectrum for the $F = \tft$ transition of W3(OH).
  See Figure \ref{w3ohi1} caption for more details.\label{w3ohi2a}}
\end{figure}

\begin{figure}
\begin{center}
\includegraphics[width=4.0in]{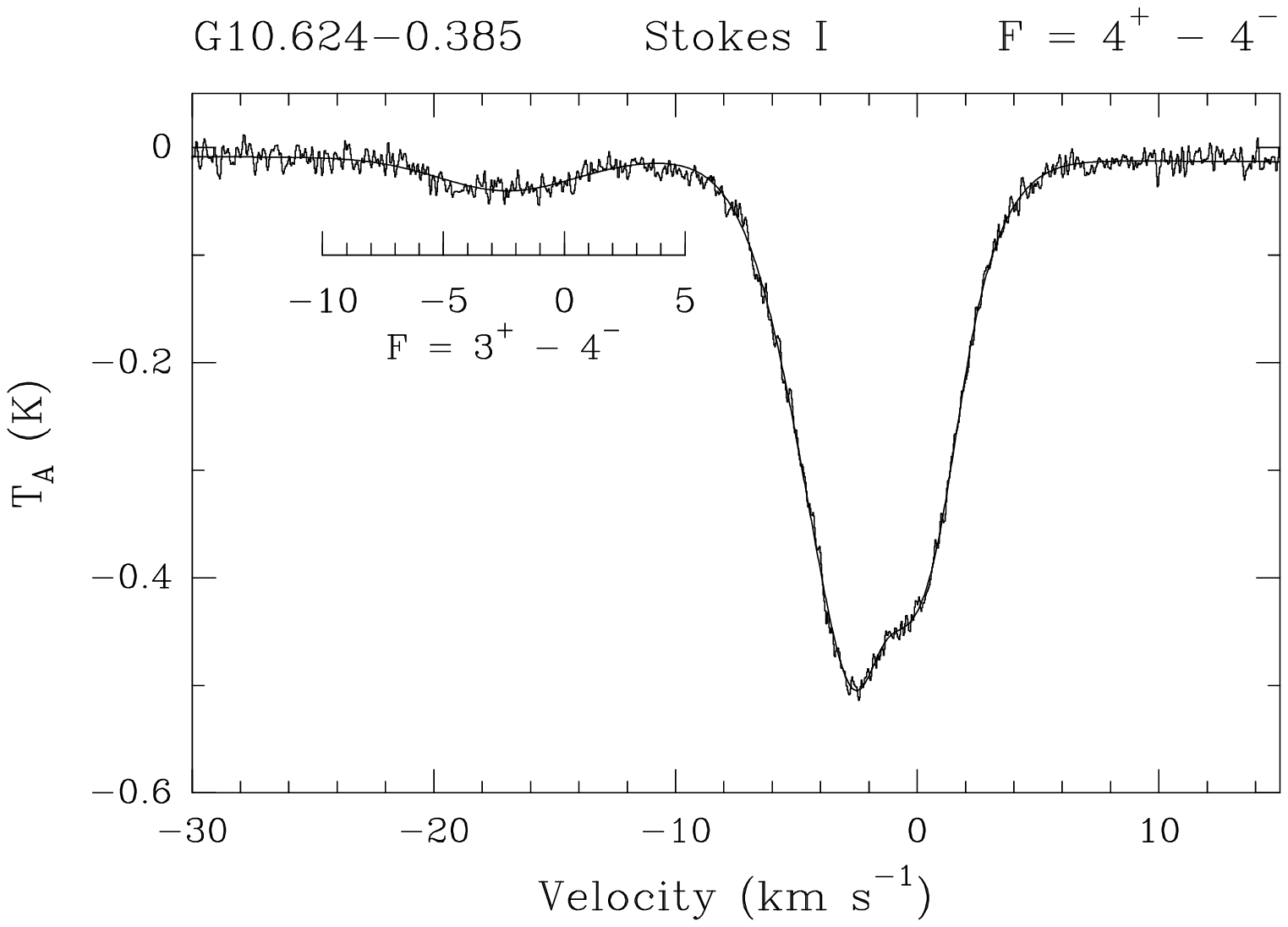}
\end{center}
\caption{Stokes I spectrum for the $F = \tff$ and $\ttf$ transitions
  of G10.624$-$0.385.  See Figure \ref{w3ohi1} caption for more
  details.  The $F = \ttf$ line is
  shown in greater detail in Figure \ref{g10i1a}.\label{g10i1}}
\end{figure}

\begin{figure}
\begin{center}
\includegraphics[width=4.0in]{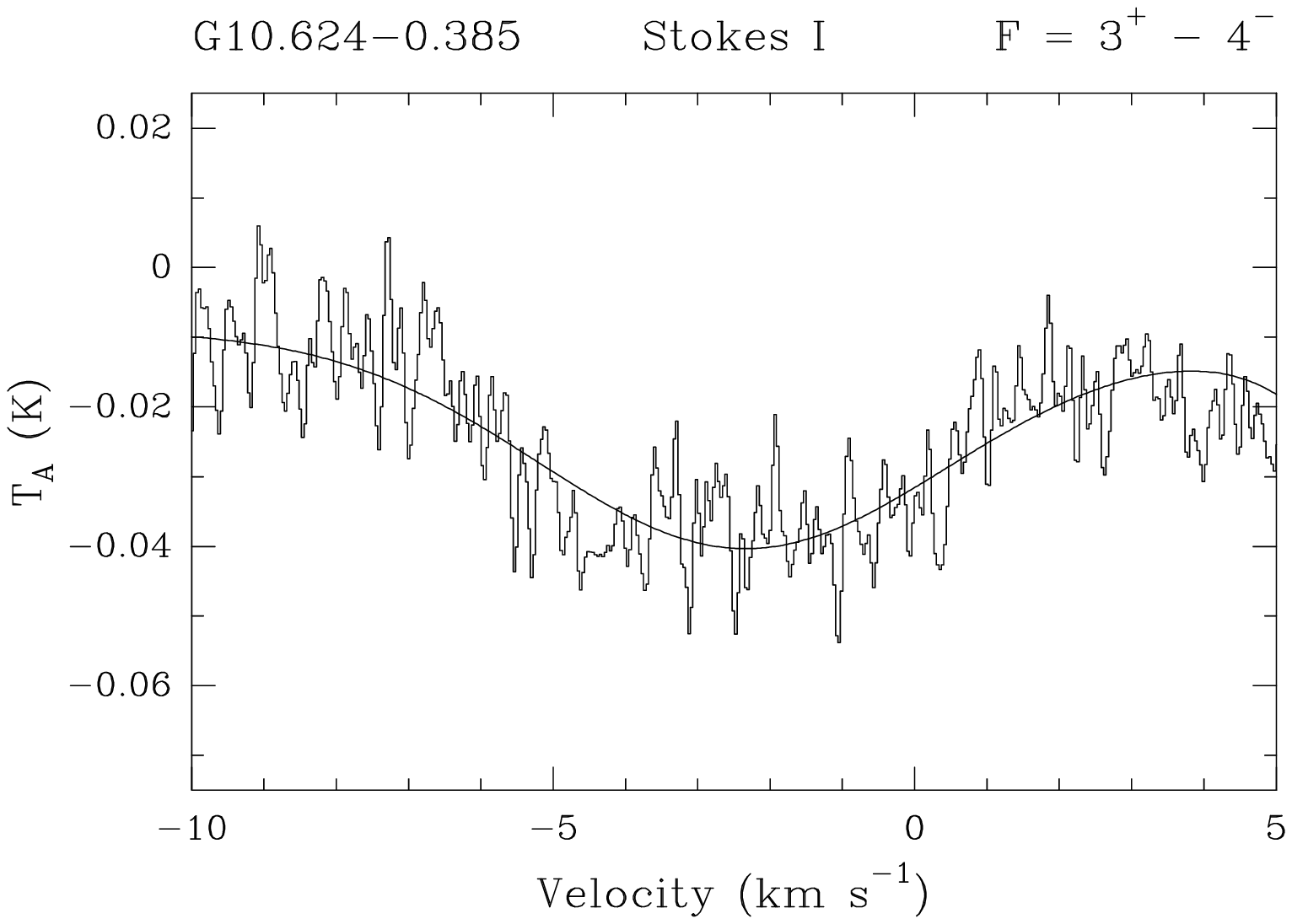}
\end{center}
\caption{Stokes I spectrum for the $F = \ttf$ transition
  of G10.624$-$0.385.  See Figure \ref{w3ohi1} caption for more
  details.  The turnoff at higher velocity is due to $F = \tff$
  absorption, shown in Figure \ref{g10i1}.\label{g10i1a}}
\end{figure}

\begin{figure}
\begin{center}
\includegraphics[width=4.0in]{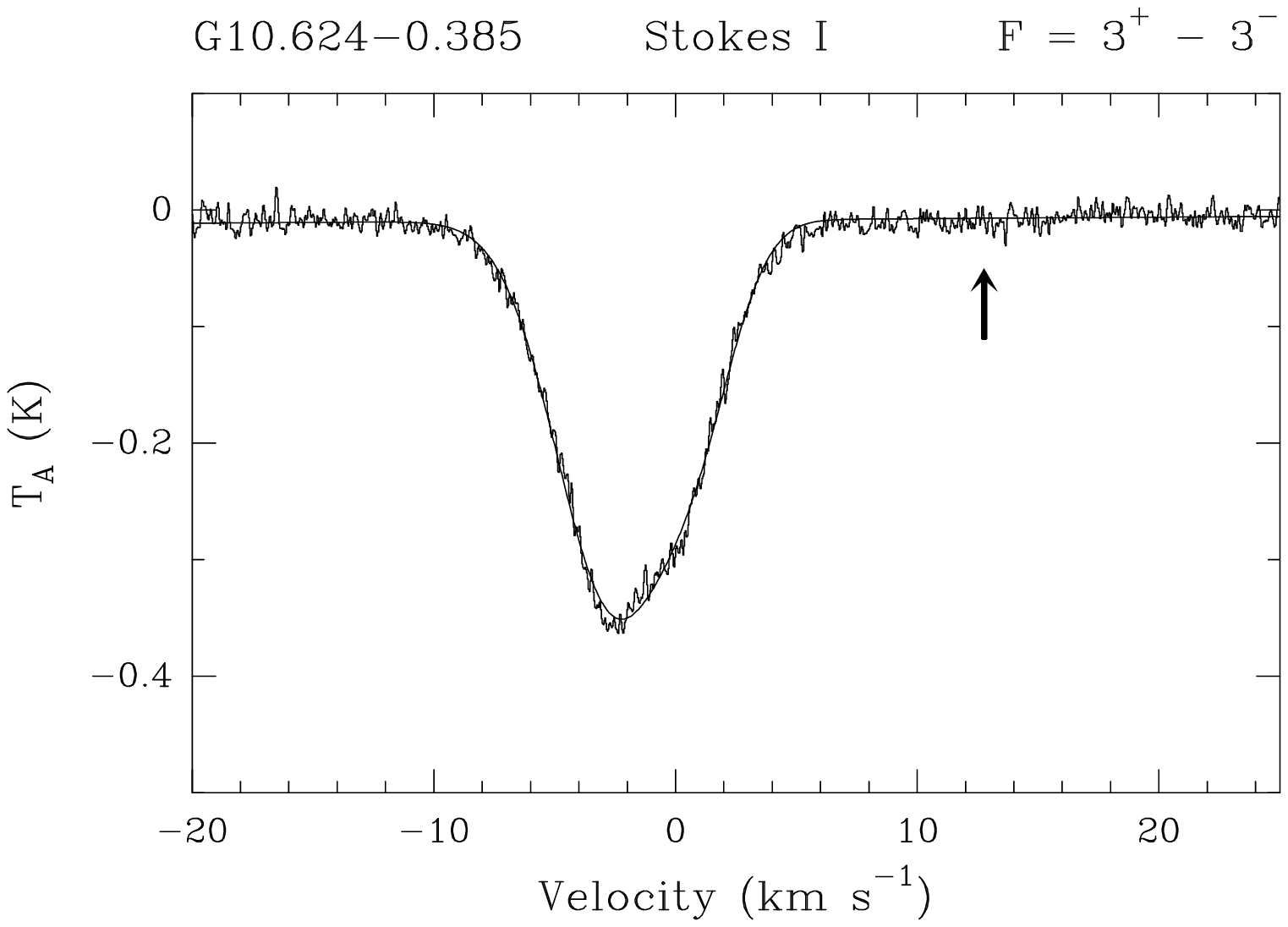}
\end{center}
\caption{Stokes I spectrum for the $F = \ttt$ transition of
  G10.624$-$0.385.  See Figure \ref{w3ohi1} caption for more details.
  The arrow indicates where $F = \tft$ absorption would appear if at
  the velocity of the main absorption component.\label{g10i2}}
\end{figure}

\begin{figure}
\begin{center}
\includegraphics[width=4.0in]{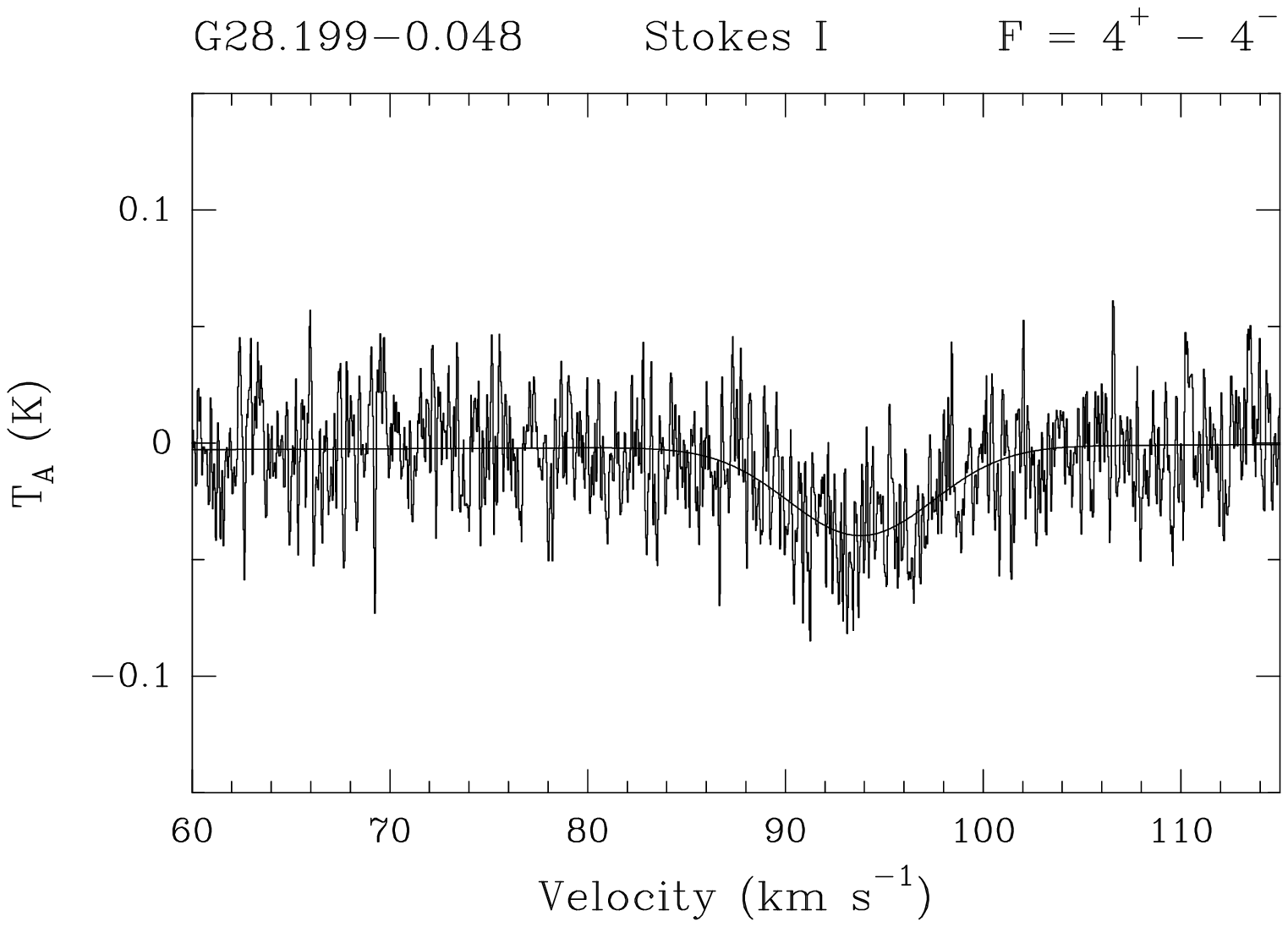}
\end{center}
\caption{Stokes I spectrum for the $F = \tff$ transition of
  G28.199$-$0.048.  See Figure \ref{w3ohi1} caption for more
  details.\label{g28i1}}
\end{figure}

\begin{figure}
\begin{center}
\includegraphics[width=4.0in]{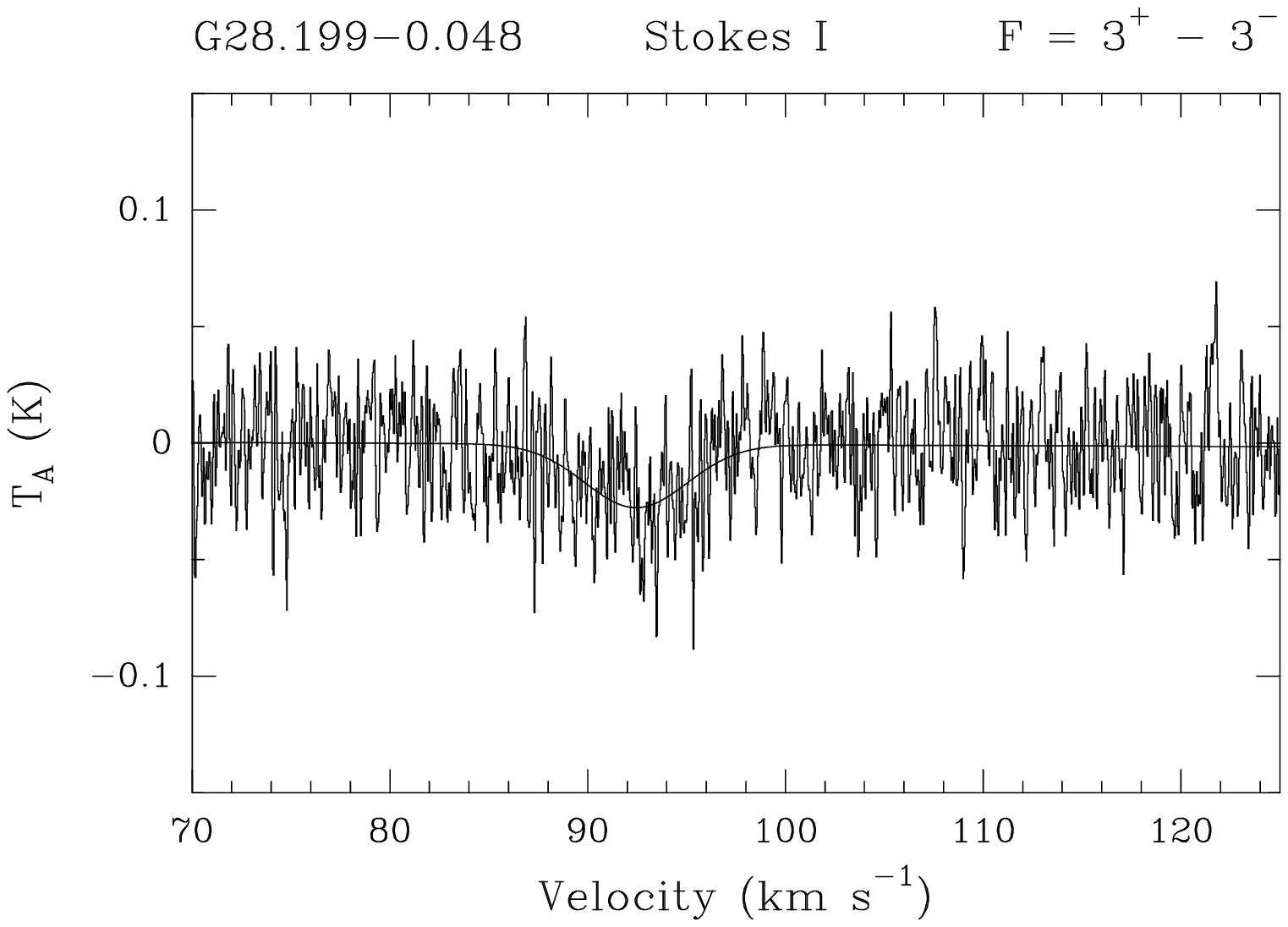}
\end{center}
\caption{Stokes I spectrum for the $F = \ttt$ transition of
  G28.199$-$0.048.  See Figure \ref{w3ohi1} caption for more
  details.\label{g28i2}}
\end{figure}

\begin{figure}
\begin{center}
\includegraphics[width=4.0in]{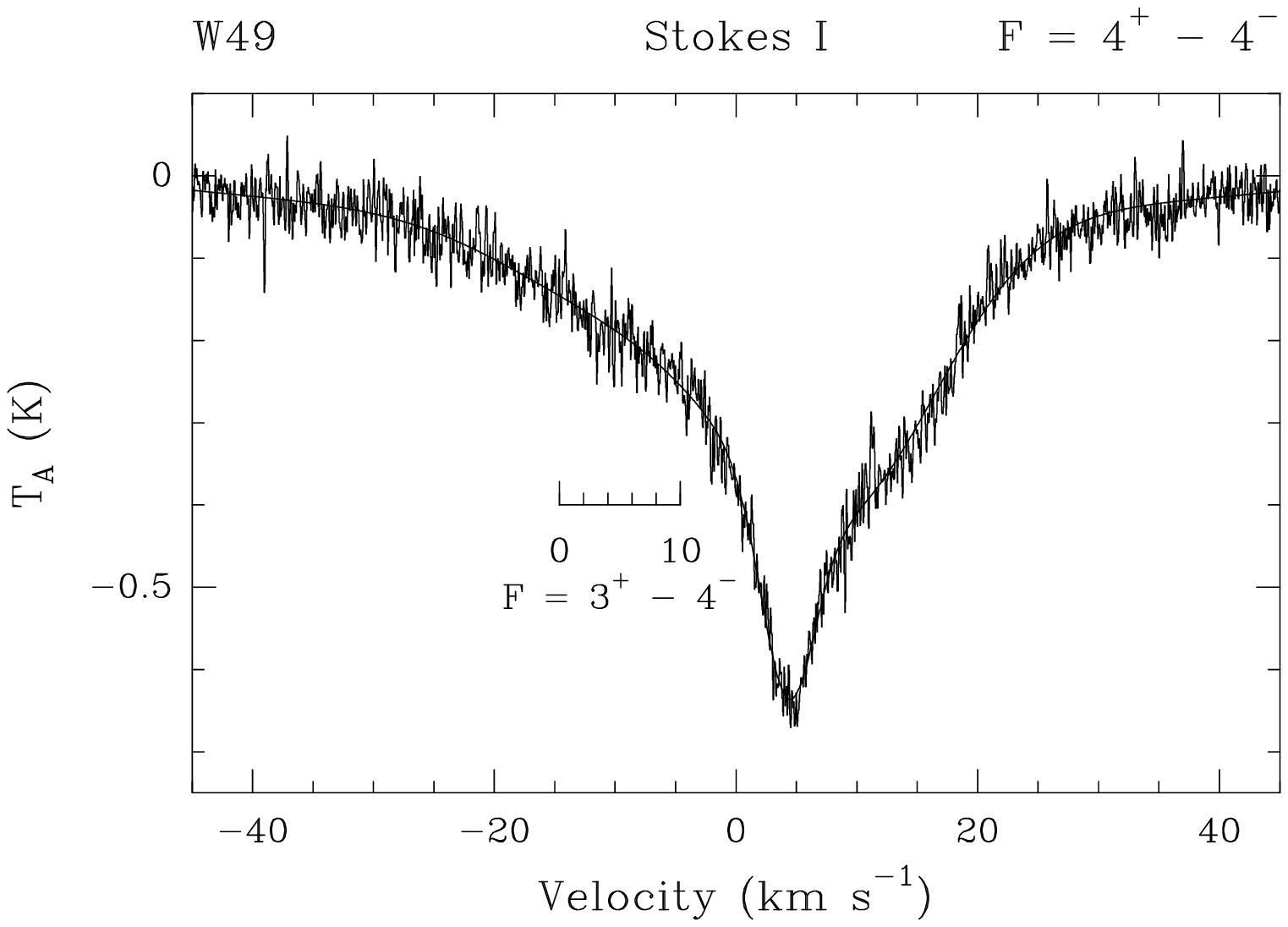}
\end{center}
\caption{Stokes I spectrum for the $F = \tff$ and $\ttf$ transitions
  of W49.  See Figure \ref{w3ohi1} caption for more
  details.\label{w49i1}}
\end{figure}

\begin{figure}
\begin{center}
\includegraphics[width=4.0in]{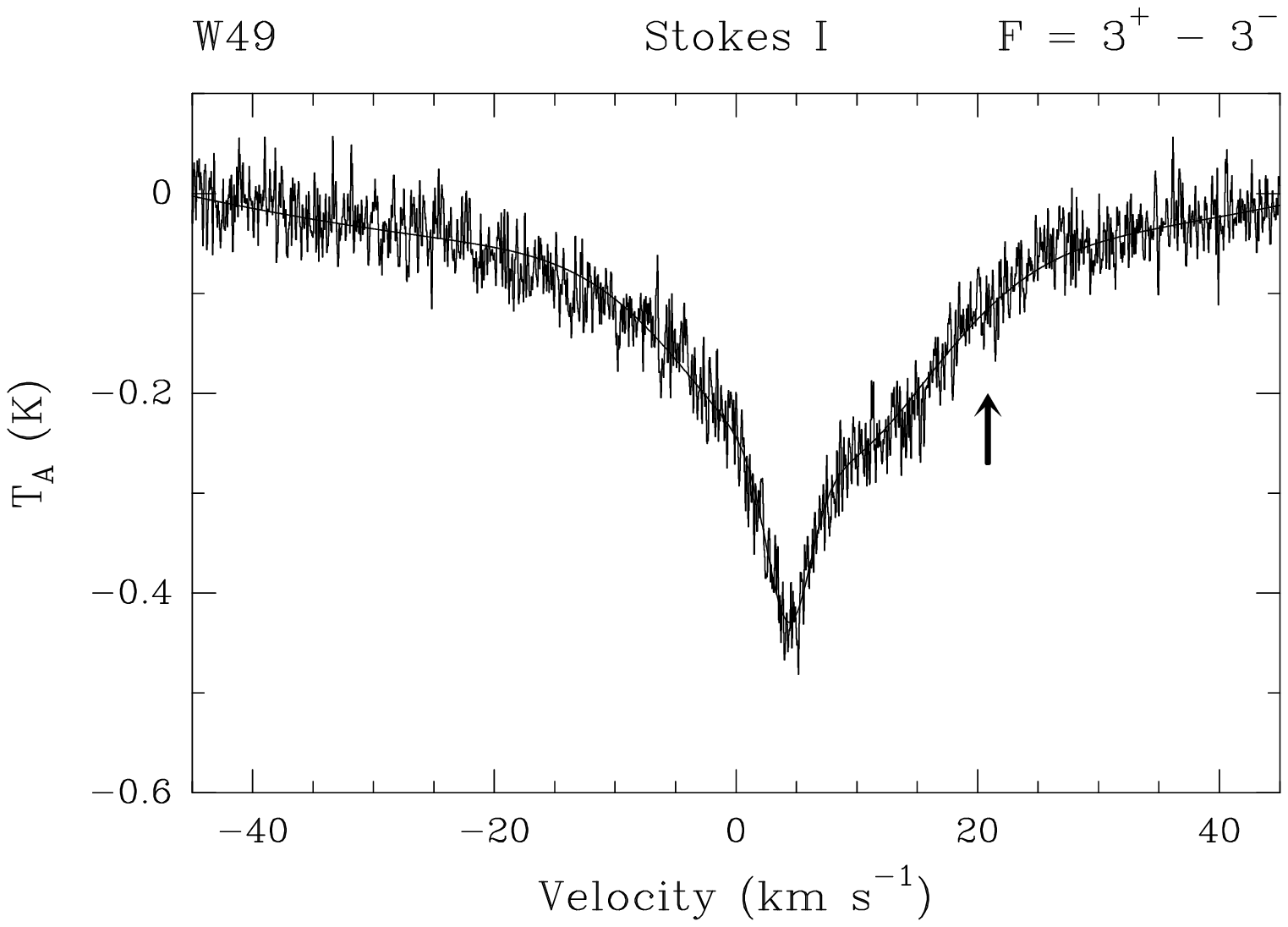}
\end{center}
\caption{Stokes I spectrum for the $F = \ttt$ transition of W49.  See
  Figure \ref{w3ohi1} caption for more details.  The arrow indicates
  where $F = \tft$ absorption would appear if at the velocity of the
  main absorption component.\label{w49i2}}
\end{figure}

\begin{figure}
\begin{center}
\includegraphics[width=4.0in]{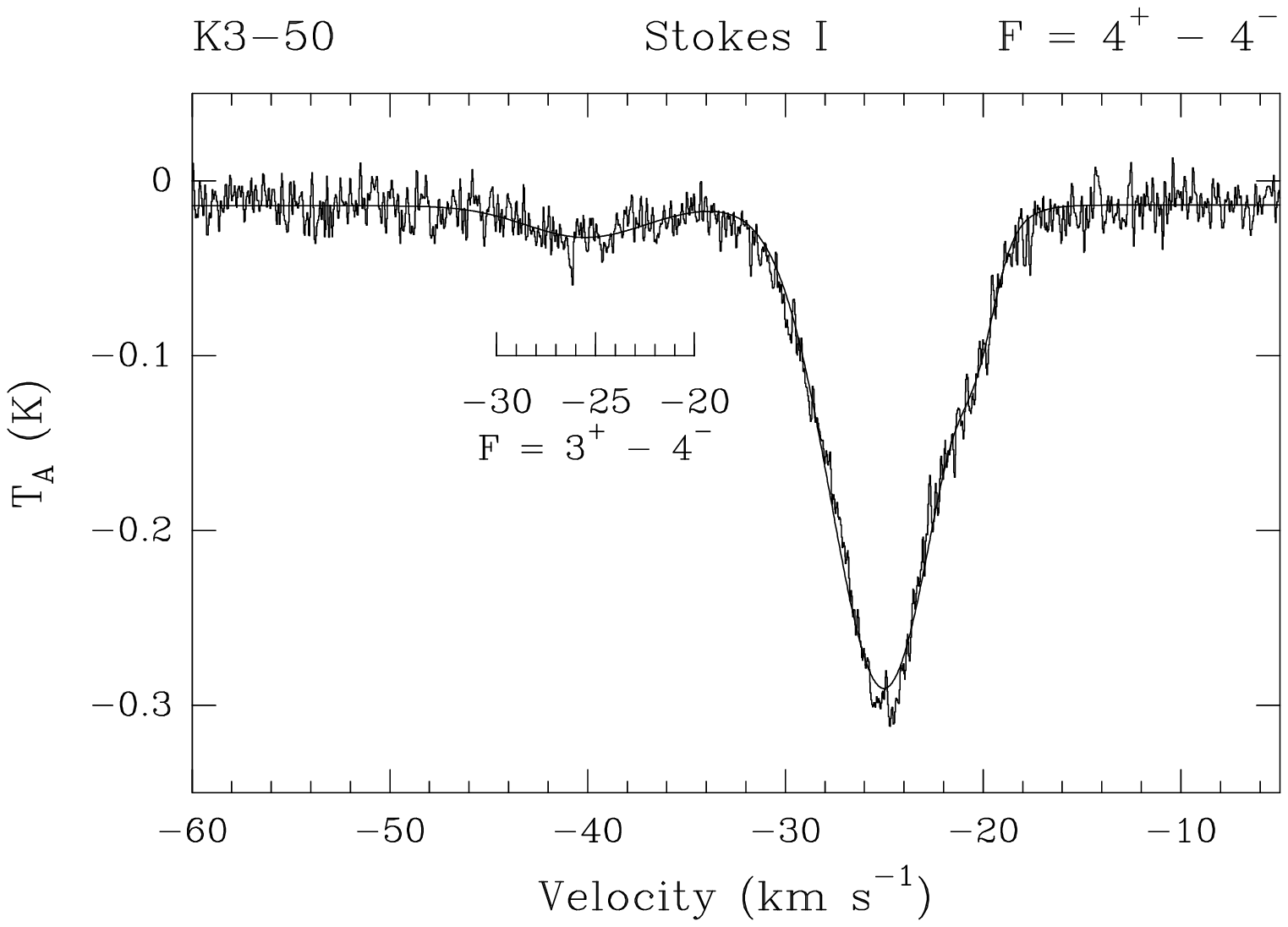}
\end{center}
\caption{Stokes I spectrum for the $F = \tff$ and $\ttf$ transitions
  of K3$-$50.  See Figure \ref{w3ohi1} caption for more
  details.  The $F = \ttf$ line is
  shown in greater detail in Figure \ref{k350i1a}.\label{k350i1}}
\end{figure}

\begin{figure}
\begin{center}
\includegraphics[width=4.0in]{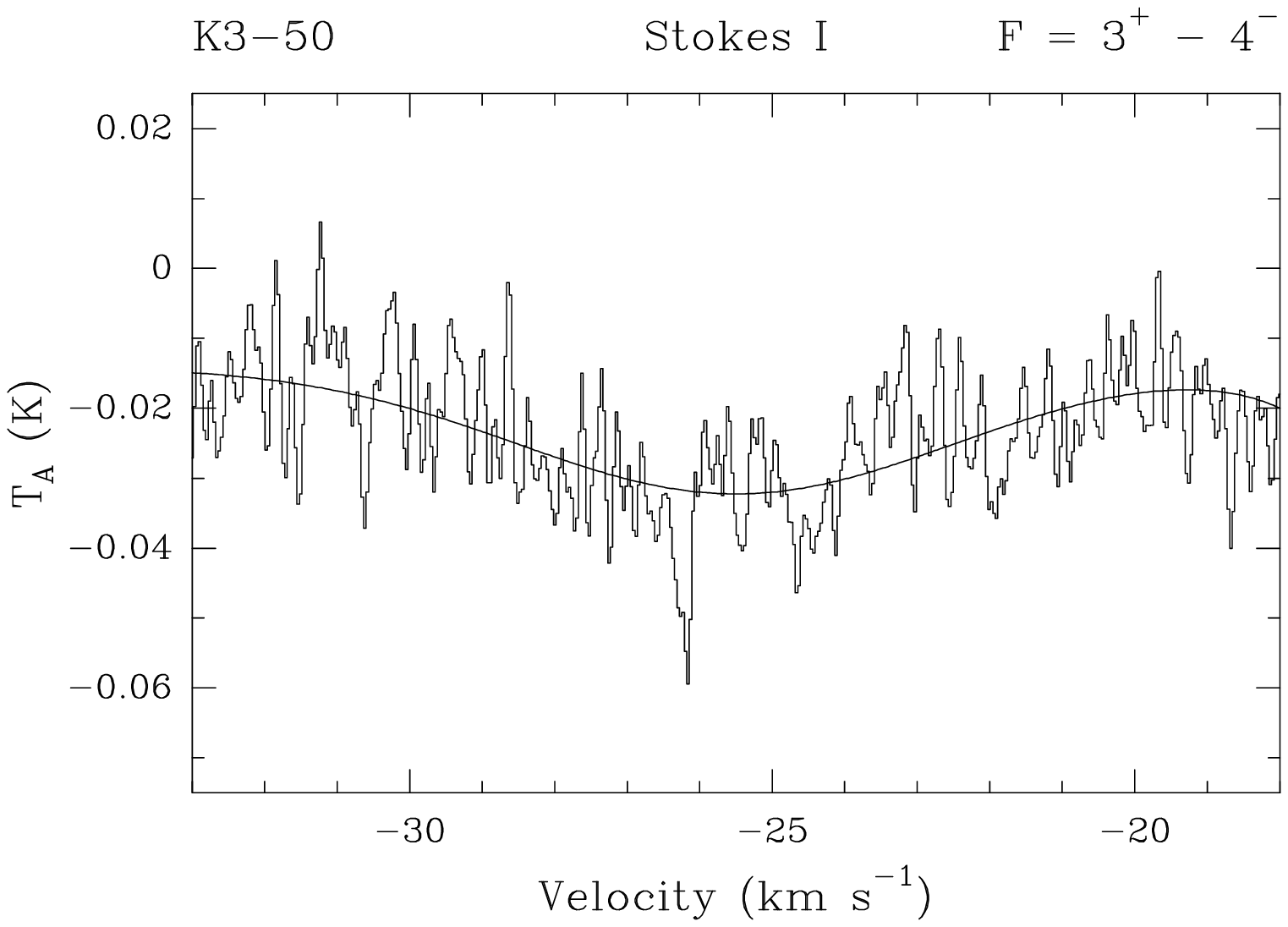}
\end{center}
\caption{Stokes I spectrum for the $F = \ttf$ transition
  of K3$-$50.  See Figure \ref{w3ohi1} caption for more
  details.  The turnoff at higher velocity is due to $F = \tff$
  absorption, shown in Figure \ref{k350i1}.\label{k350i1a}}
\end{figure}

\begin{figure}
\begin{center}
\includegraphics[width=4.0in]{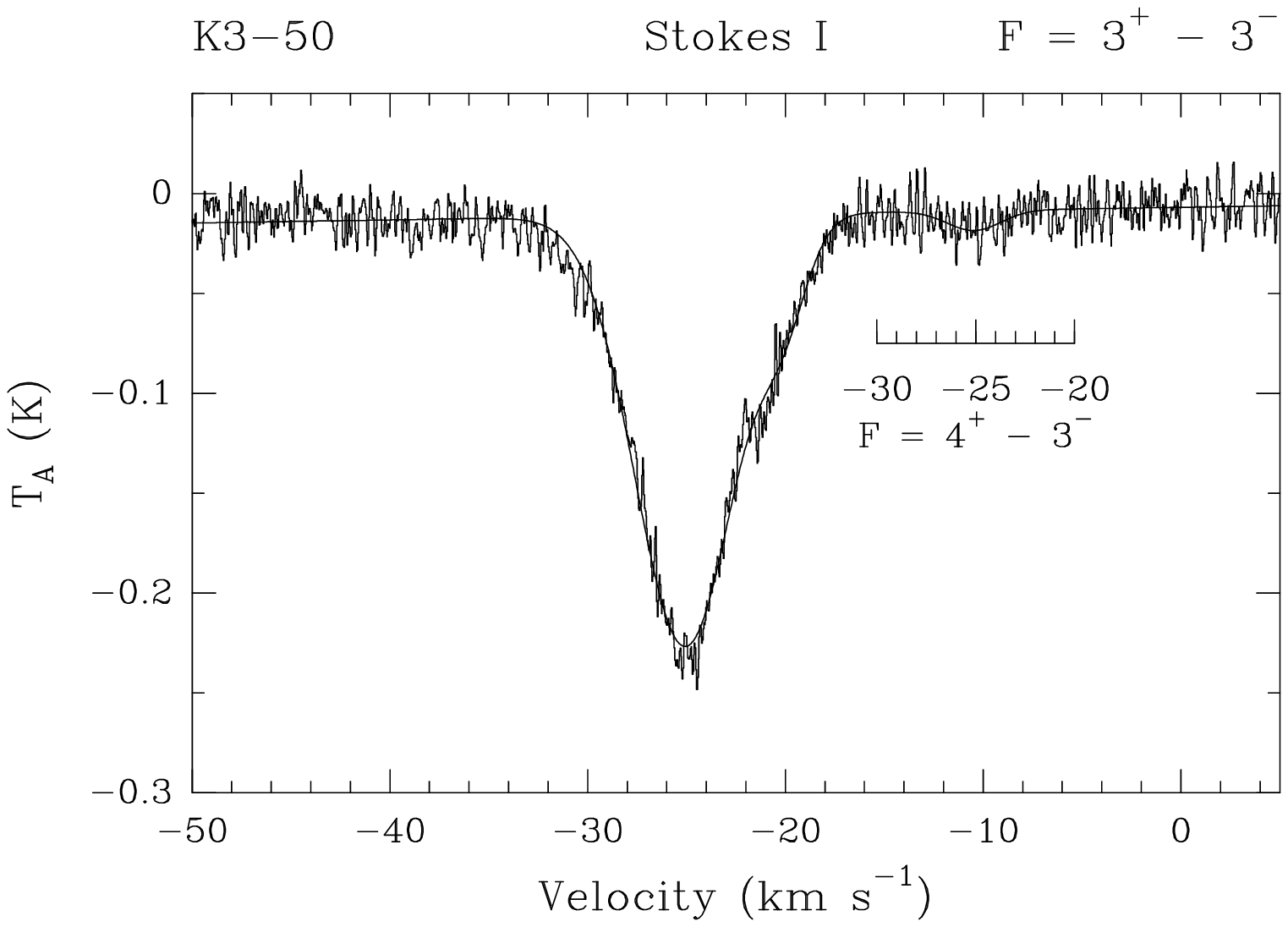}
\end{center}
\caption{Stokes I spectrum for the $F = \ttt$ and $\tft$ transitions
  of K3$-$50.  See Figure \ref{w3ohi1} caption for more
  details.  The $F = \tft$ line is
  shown in greater detail in Figure \ref{k350i2a}.\label{k350i2}}
\end{figure}

\begin{figure}
\begin{center}
\includegraphics[width=4.0in]{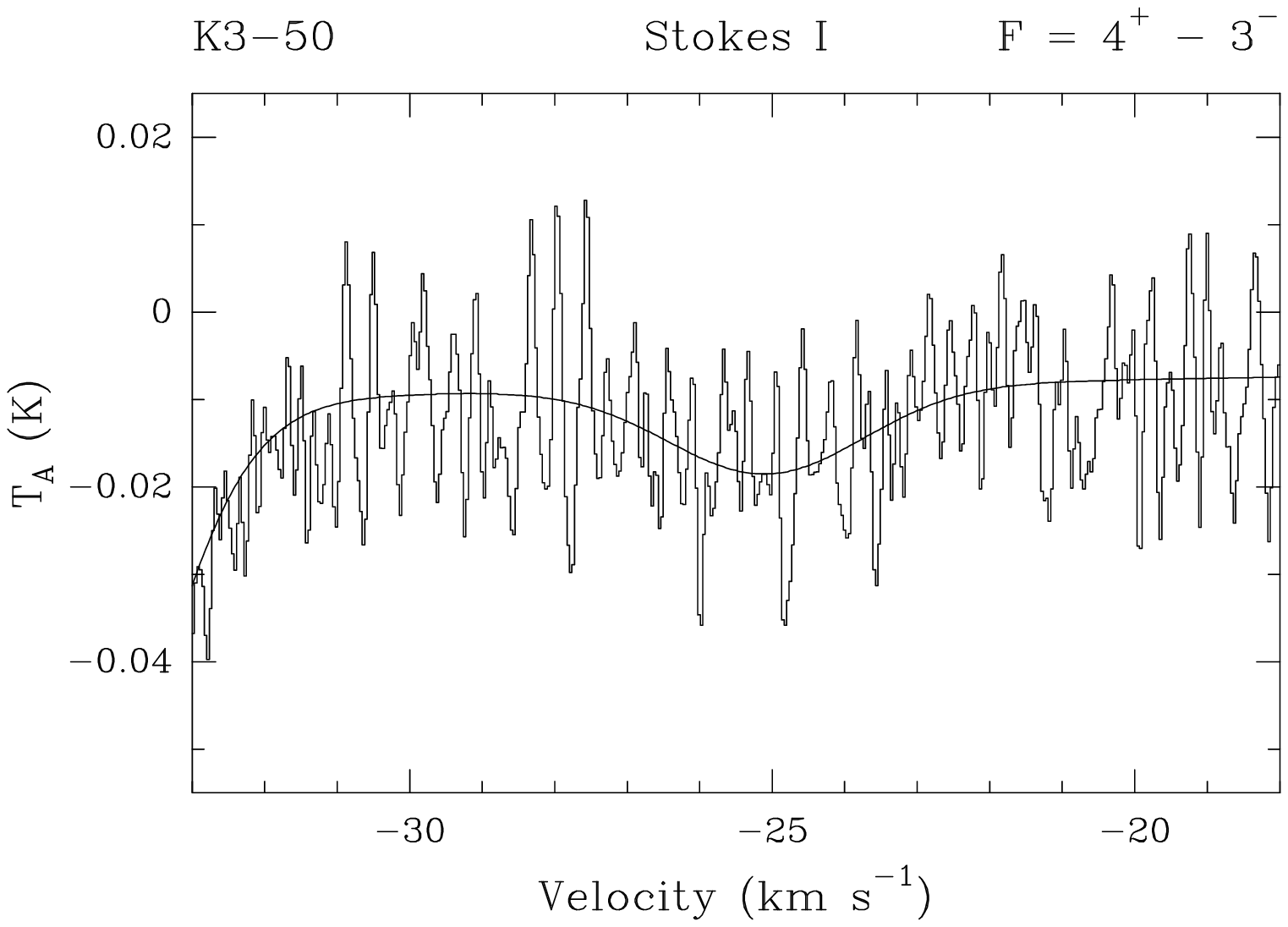}
\end{center}
\caption{Stokes I spectrum for the $F = \tft$ transition
  of K3$-$50.  See Figure \ref{w3ohi1} caption for more
  details.  The turnoff at lower velocity is due to $F = \ttt$
  absorption, shown in Figure \ref{k350i2}.\label{k350i2a}}
\end{figure}

\begin{figure}
\begin{center}
\includegraphics[width=4.0in]{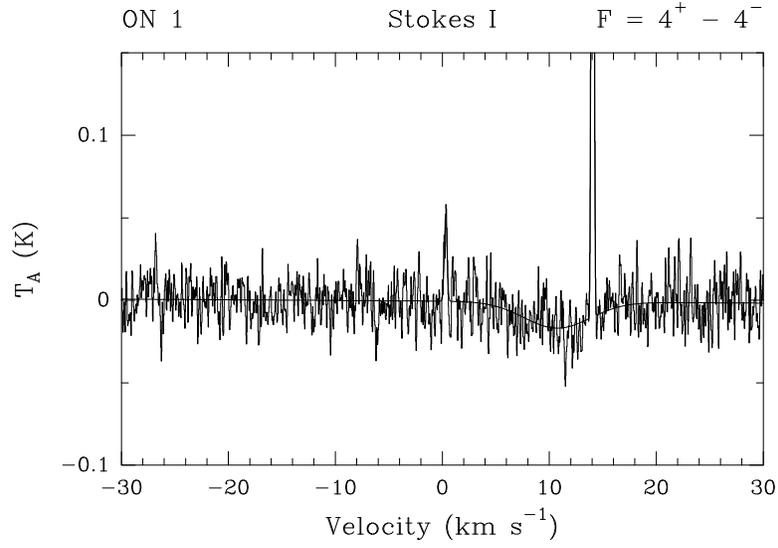}
\end{center}
\caption{Stokes I spectrum for the $F = \tff$ transition of ON~1.  See
  Figure \ref{w3ohi1} caption for more details.  The masers at $0$ and
  $14$~\kms\ are plotted in Figures \ref{on1lr1} and
  \ref{on1lr1a}.\label{on1i1}}
\end{figure}

\begin{figure}
\begin{center}
\includegraphics[width=4.0in]{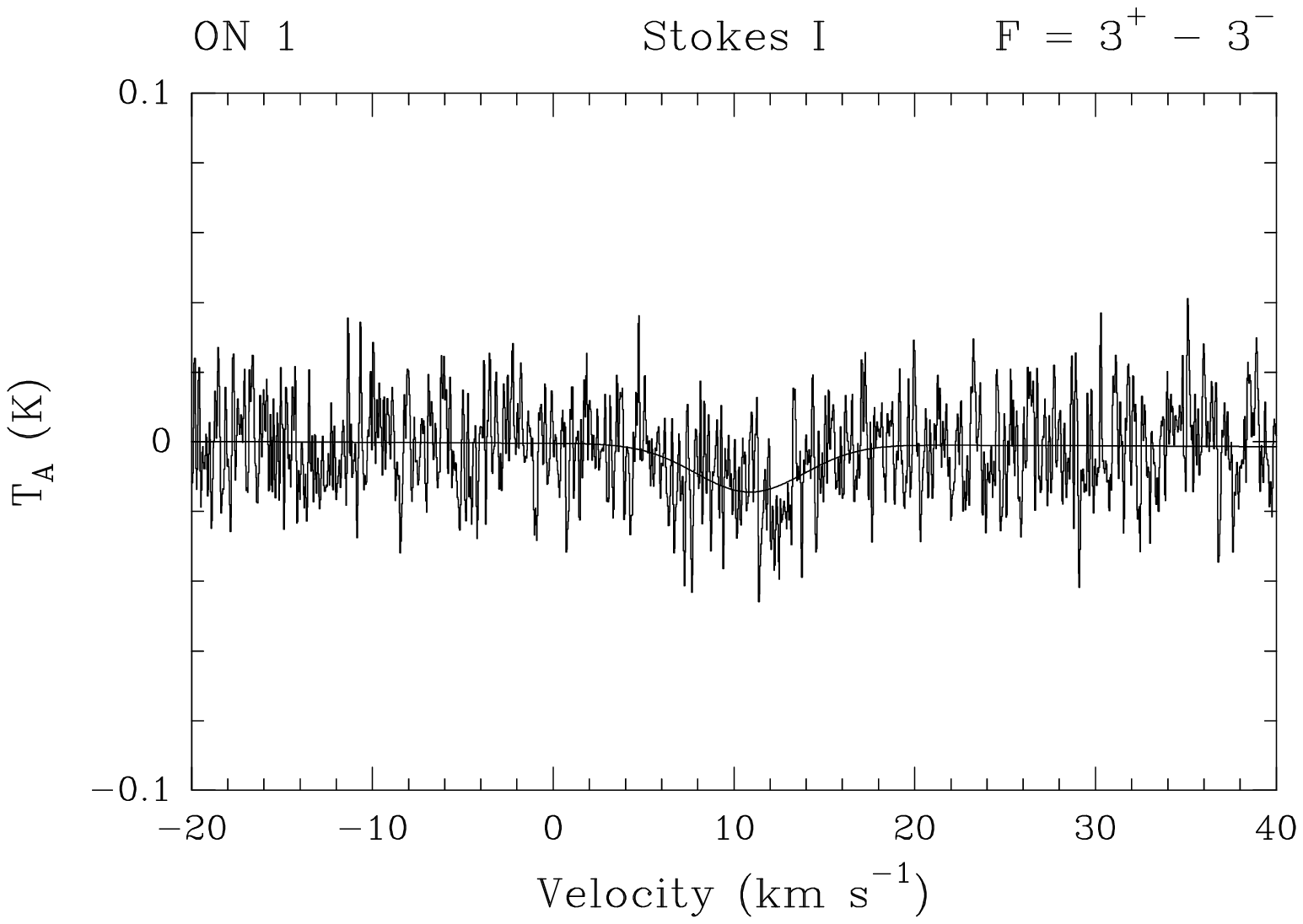}
\end{center}
\caption{Stokes I spectrum for the $F = \tff$ transition of ON~1.  See
  Figure \ref{w3ohi1} caption for more details.\label{on1i2}}
\end{figure}

The masers we detected are shown in Figures \ref{w3ohlr1} to
\ref{on1lr1a} and discussed in \S \ref{masers}.  The absorption
features we detected are shown in Figures \ref{w3ohi1} to \ref{on1i2}
and discussed in \S \ref{absorption}.  The absorption spectra have
been Hanning weighted for clarity, although the analyses are based on
the uniform-weighted data.  The antenna temperatures
are related to the flux density by $S = 2 k T_A/A$, where $A$ is the
effective collecting area of the telescope.  We have used the
convention that Stokes $\mathrm{I} = 0.5 \, (T_{A,\mathrm{LCP}} +
T_{A,\mathrm{RCP}})$ and Stokes $\mathrm{V} = 0.5 \,
(T_{A,\mathrm{LCP}} - T_{A,\mathrm{RCP}}).$\footnote{We define Stokes
V as LCP $-$ RCP to be consistent with \citet{gusten}.}  Thus, $T_A =
1$~K in both RCP and LCP would result in $S = 0.59$~Jy in Stokes I.
Zeeman
measurements in the OH absorption are discussed in \S \ref{zeeman}.
Remarks about absorption line ratios are presented in \S
\ref{line-ratios}.

\subsection{Maser Emission\label{masers}}

Masers were found in two sources: W3(OH) and ON~1.  Parameters of the
maser lines are listed in Table \ref{maser-table}.  Magnetic field
estimates assume a Zeeman splitting coefficient of 0.178
km~s$^{-1}$~mG$^{-1}$ for the \tff transition and 0.230
km~s$^{-1}$~mG$^{-1}$ for the \ttt transition \citep{gusten}.  Zeeman
splitting of OH masers is sensitive to the strength of the \emph{full,
three-dimensional} magnetic field, independent of its inclination to
the line of sight, when the Zeeman splitting exceeds the linewidth.

In W3(OH), we find maser lines in both the \sevenhalves, $F = \tff$
and $\ttt$ transitions.  We detect seven or eight maser line components in each
circular polarization in the $F = \tff$ transition, compared to the
three previously detected with other single dish antennas
\citep{baudry81,gusten,baudry02}.  Nearly all of these lines can be
grouped into Zeeman pairs with implied full magnetic field strength
ranging from 6.9 to 11.3 mG.  This range agrees with previous
observations by \citet{baudry98} with the VLBA, whose angular
resolution is sufficient to unambiguously pair most Zeeman components.
The strongest masers we detect have counterparts in Table 1 of Baudry
\& Diamond.  It is not possible to identify unambiguous counterparts
to our weaker masers, which may be blends of features resolved at the
submilliarcsecond resolution of the VLBA.  The detection of the LCP
and RCP masers centered at $-44.85$~\kms\ is new.  There is some
ambiguity as to whether the LCP maser at $-43.66$~\kms\ should be
paired with the RCP maser at $-43.54$ or $-43.51$~\kms, but the former
is more likely because the linewidth of the latter is much greater
than for the LCP maser.  In the $F = \ttt$ transition, we find one
Zeeman pair implying a line-of-sight magnetic field component of 10.3
mG, consistent with the detection by \citet{gusten}.

ON~1 is the only other source in which we find \sevenhalves\ masers.
The $F = \tff$ maser at 14~\kms\ was also detected by
\citet{baudry02}, although their velocity resolution was insufficient
to find measurable Zeeman splitting.  We do not see the strong (0.5
Jy) maser at $-$0.13~\kms\ that they do; however, we find evidence of
a weak maser at $0.38$~\kms\ (LCP) and $0.24$~\kms\ (RCP).  While
these lines are weak, we believe they are real.  The peaks of the LCP
and RCP lines are only $3.8$ and $2.8$ times the single-channel RMS
noise of 0.02~K, respectively, but the line widths are 8 times a
single channel width.  These masers can be interpreted either as $F =
\tff$ lines at 0.3~\kms\ or as $F = \ttf$ lines at 14.9~\kms.  The
1665 MHz (\threehalves) masers in ON~1 are grouped in two
velocity ranges: 1 to 4~\kms\ and 10 to 16~\kms\ \citep{arm};
likewise, the 6035 MHz (\fivehalves) masers are seen from $-1$
to 2~\kms\ and 13 to 16~\kms\ \citep{baudry97}.  Thus, either the $F
= \tff$ or $F = \ttf$ interpretation of these maser lines is
consistent with the ground-state OH maser velocities.

Nevertheless, two arguments suggest that the weak maser feature in
ON~1 is a maser in the $F = \tff$ transition.  First, satellite-line
emission in interstellar OH masers is generally weak compared to
main-line masers.  This is true in the \threehalves\ lines, where the
1665 and 1667 MHz masers are typically much stronger than 1612 and
1720 MHz masers.  In the \fivehalves\ lines, strong masers are found
in the 6035 and 6030 MHz main lines, but 6049 MHz satellite-line
emission is weak and rare \citep{baudry97}, while the 6016 MHz
satellite line has not been seen in emission
\citep{baudry97,gardner83}.  Indeed, if photon trapping is important,
inversion in the 6016 MHz satellite line may be impossible
\citep{elitzur77}.  As for the \sevenhalves\ lines, while about a
dozen masers have been found in the 13441 MHz main line transition,
the 13434 MHz maser in W3(OH) remains the single known maser in the
other main line transition, despite two searches encompassing 77
distinct interstellar masing sources \citep{baudry02,caswell04}.
Empirically, this suggests that satellite-line \sevenhalves\ masers
should be extremely rare if they exist at all.  Second, due to a
smaller Zeeman coefficient, the magnetic field strength implied by a
constant-velocity Zeeman splitting is higher for satellite lines than
for main lines.  If the weak maser in ON~1 is actually in the $F =
\ttf$ transition, the implied magnetic field is $-14.5$~mG.  This
field strength would be much larger than that derived from any other
Zeeman pair in ON~1 in the $^2\Pi_{3/2}$ ladder \citep[see
also][]{baudry97, fishthesis}.

\begin{deluxetable}{lllllllll}
\tablewidth{0 pt}
%

%
%

\newlength{\minuswidth}
\newlength{\numberwidth}
\settowidth{\minuswidth}{$-$}
\settowidth{\numberwidth}{$8$}
\addtolength{\minuswidth}{-\numberwidth}
\newcommand{\phmn}{\hspace*{\minuswidth}}


\tablecaption{Maser Parameters\label{maser-table}}
\tablehead{
            \multicolumn{2}{c}{} &
            \multicolumn{3}{c}{\hrulefill LCP\hrulefill} &
            \multicolumn{3}{c}{\hrulefill RCP\hrulefill} &
            \colhead{} \\
            \colhead{Source} &
            \colhead{Transition} &
            \colhead{$v\lsr$} &
            \colhead{$T_A$} &
            \colhead{$\Delta v$} &
            \colhead{$v\lsr$} &
            \colhead{$T_A$} &
            \colhead{$\Delta v$} &
            \colhead{$B$} \\
            \colhead{} &
            \colhead{} &
            \colhead{(\kms)} &
            \colhead{(K)} &
            \colhead{(\kms)} &
            \colhead{(\kms)} &
            \colhead{(K)} &
            \colhead{(\kms)} &
            \colhead{(mG)}
          }
\startdata
W3(OH) & \ttt& $-$42.53 &\phn0.15 & 0.32 & $-$42.29 &\phn0.15 & 0.29 &\phmn10.3 \\
       & \tff& $-$44.94 &\phn1.93 & 0.18 & $-$44.75 &\phn1.82 & 0.20 &\phmn10.5 \\
       &     & $-$43.97 &\phn4.01 & 0.38 & $-$43.84 &\phn0.56 & 0.22 &\phmin7.0 \\
       &     & $-$43.66 &\phn9.77 & 0.24 & $-$43.54 &\phn7.02 & 0.19 &\phmin6.9 \\
       &     &          &         &      & $-$43.51 &\phn6.22 & 0.70 & \nodata\tnm{a}\\
       &     & $-$43.33 &   12.31 & 0.15 & $-$43.19 &   12.36 & 0.17 &\phmin8.2 \\
       &     & $-$43.08 & 62.65   & 0.30 & $-$42.93 & 63.27   & 0.31 &\phmin8.5 \\
       &     & $-$42.46 & 64.50   & 0.27 & $-$42.26 & 63.35   & 0.28 &\phmn11.3 \\
ON 1   &\tff &\phmin14.09&\phn1.11 &0.20&\phmin14.02&\phn1.14& 0.21 &$-$3.8\\
       &     &\phmin\phn0.38\tnm{b}&\phn0.07 & 0.26&\phmin\phn0.24\tnm{b} &\phn0.06 & 0.30 & $-$8.3\tnm{b}\\
\enddata
\tnt{a}{Magnetic field strength is 8.5~mG if paired with LCP maser at
  $-43.66$~\kms.  See \S \ref{masers} for details.}
\tnt{b}{Table velocities and magnetic field assume that these
are $F = \tff$ masers.  See \S \ref{masers} for
details.}
\end{deluxetable}

\subsection{Absorption\label{absorption}}

Unlike maser emission, absorption is clearly seen in all sources in
our sample in both the $F = \ttt$ and $F = \tff$ transitions.  In some
sources, absorption is seen in the $F = \ttf$ and/or $F = \tft$
transitions as well.  We believe that this is the first clear
detection of absorption in the \sevenhalves\ satellite lines.
Parameters of the absorption lines are given in Table
\ref{absorption-table}.  We summarize results for individual sources
below.

W3(OH): Absorption is seen in all four hyperfine transitions in
W3(OH).  The presence of multiple strong maser lines (see Figures
\ref{w3ohlr1} and \ref{w3ohi1}) in the $F = \tff$ transition is
problematic for fitting Gaussian components to the absorption.  The
data in the velocity range from $-45.52$ to $-41.00$~\kms\ have been
excluded from the fitting of the two Gaussian components listed in
Table \ref{absorption-table}.

G10.624$-$0.385: Absorption is seen in the $F = \ttf$ transition as
well as the main lines.

G28.199$-$0.048: Absorption is seen in the main lines only.  The noise
is high due to the short time spent on source.

W49: Very broad absorption is seen in the main lines.  This absorption
is broader than the frequency difference between adjacent main- and
satellite-line transitions.  It is possible that the absorption marked
at $5.59$~\kms\ in the $F = \ttf$ transition is actually at
$-9.01$~\kms\ in the $F = \tff$ transition.  However, the velocity
assuming that this component is due to the $F = \ttf$ transition is
more consistent both with the other two transitions and with the
H92$\alpha$/He92$\alpha$ velocity of $6.2 \pm 2.3$\kms\ for the
associated \hii\ region \citep{depree97}.

K3$-$50: Absorption is seen in all four hyperfine transitions.

ON 1: Absorption is seen in the main lines only.  While the
observations occurred during a period devoid of precipitation, the
noise is still high due to the short time spent on source.

\subsection{Zeeman Measurements in Absorption\label{zeeman}}

The magnetic field can be measured by Zeeman splitting in absorption,
as well as in maser emission.  While the full three-dimensional
magnetic field strength is obtained from Zeeman splitting of
velocity-separated maser components, absorption lines are much broader
than the velocity separation between LCP and RCP components, so only
the line-of-sight component of the magnetic field can be measured
\citep[e.g.,][]{sault90}.
The Stokes V curve is related to the derivative of the Stokes I curve
by the following equation:
\begin{equation}
T_\mathrm{V}(\nu) = -C \,
  \frac{dT_\mathrm{I}(\nu)}{d\nu}
  \mathrm{B}_\parallel, \label{deriv-equation}
\end{equation}
where $T_\mathrm{V}(\nu)$ and $T_\mathrm{I}(\nu)$ are the brightness
temperatures of the Stokes V and I spectra, and $C = -1.06 \times
10^6$~Hz~G$^{-1}$ for the $F = \ttt$ transition and $7.95 \times
10^5$~Hz~G$^{-1}$ for the $\tff$ transition \citep{gusten}.
Hence, if
the Stokes V spectrum of an absorption component is nonzero, the
line-of-sight magnetic field can be measured.

\begin{figure}
\begin{center}
\includegraphics[width=4.0in]{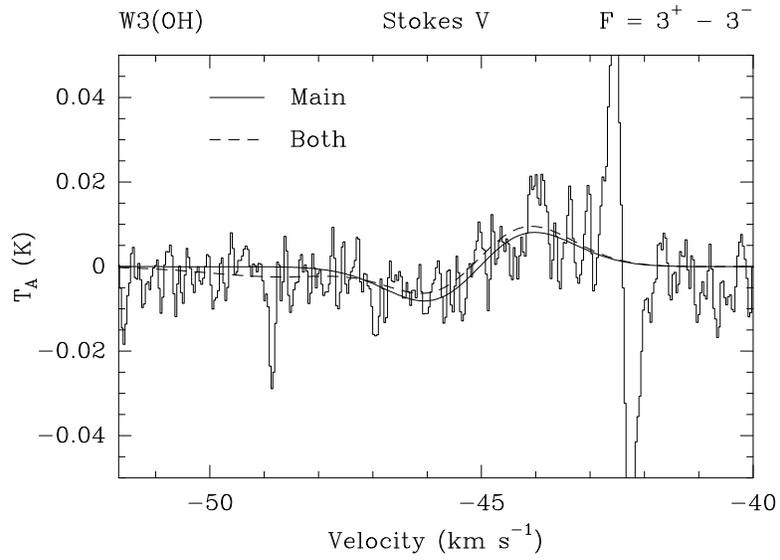}
\end{center}
\caption{Stokes V spectrum for the $F = \ttt$ transition of W3(OH).
  The solid line shows the best fit for the derivative of Stokes I for
  the main absorption feature at $-45$~\kms, and the dashed line shows
  the best fit for both $F = \ttt$ absorption features.  The fit lines
  correspond to a magnetic field of $+$3.0~mG, as explained in \S
  \ref{zeeman}.  The large value of Stokes V near $-42.4$~\kms\ is due
  to a pair of maser features, which have been excluded from the
  fit.\label{w3ohv2}}
\end{figure}

W3(OH) is the only source in which the Stokes V absorption spectrum
shows clear evidence of Zeeman splitting.  Figure \ref{w3ohv2} shows
the Stokes V spectrum of the $F = \ttt$ transition of W3(OH).  The
feature near $-42.4$~\kms\ is due to the maser.  In the direction of
lower velocity from this feature is an ``S-curve.''  Superposed atop
the data are two curves corresponding to the derivative of the
Gaussian fits to the Stokes I curve.  The curve labelled ``Main''
indicates the scaled derivative of the Gaussian centered at
$-45.03$~\kms\ (listed in Table \ref{absorption-table}), while the
curve labelled ``Both'' indicates the scaled derivative of both this
Gaussian and the one centered at $-47.22$~\kms.  A positive magnetic
field shifts RCP to higher velocity than LCP; in \emph{absorption}, this
corresponds to the positive bump of the Stokes V curve being at higher
velocity than the negative bump.  Note that in \emph{emission}, a
positive magnetic field results in the positive bump of the Stokes V
curve being at \emph{lower velocity} than the negative bump, as is
seen for the 42.4~\kms\ maser Zeeman pair in Figure \ref{w3ohv2}.
Thus the absorption and emission magnetic fields in W3(OH) are
consistent in sign even though the S-curves appear inverted with
respect to each other.

The magnetic field
value that provides the best fit to the data is $2.9 \pm 0.3$~mG for
the main feature and $3.0 \pm 0.3$~mG for both features combined.
This is consistent with the values of $3.1 \pm 0.4$~mG and $3.2 \pm
0.6$~mG obtained by \citep{gusten}.

For the other sources, no S-curve is observed in the Stokes V
spectra, but we can place upper limits on the possible magnetic field
strength.  From Equation \ref{deriv-equation}, the measurement error
of the line-of-sight magnetic field component in a single channel is
\begin{equation}
\sigma_{B_\parallel} =
  \frac{-\sigma_{T_\mathrm{V}}}{C \frac{dT_\mathrm{I}}{d\nu}}.
\label{error-equation}
\end{equation}
In principle, each spectral channel constitutes an independent
measurement of $\sigma_{B_\parallel}$.  The overall error of the
parallel magnetic field strength can be estimated as
\begin{equation}
\sigma_{B_\parallel,\mathrm{overall}} =
  \left( \sum \sigma^{-2}_{B_\parallel} \right)^{-1/2},
\end{equation}
where the summation is taken over all spectral channels.

Table \ref{zeeman-table} gives the $3\sigma$ upper limits on magnetic
field strengths from absorption measurements.  For comparison, the
average three-dimensional magnetic field strengths obtained from OH
maser Zeeman splitting are also provided.  Masers from each of the $J
= 3/2$, $5/2$, and $7/2$ sets of transitions are considered
separately.  When Zeeman splitting is detected in more than one set of
transitions, the minimum and maximum average magnetic strengths
obtained from OH maser Zeeman splitting in sets of transitions
corresponding to different values of $J$ are quoted.  The actual range
of magnetic field strengths seen in maser Zeeman splitting may be
greater that that listed in Table \ref{zeeman-table}, but averages
seem appropriate for comparison with our absorption data, since the
GBT beamwidth is much greater than the angular extent of any of our
sources.  With the exception of the Zeeman pair centered at
$-44.85$~km~s$^{-1}$, Zeeman pairs in the $F = \tff$ transition
detected in this work are excluded from the calculation of the average
magnetic field strength in W3(OH) obtained from $J = 7/2$ masers, in
deference to the much higher angular resolution obtained by
\citet{baudry98}.  Note that the magnetic fields derived from the
Zeeman pairs we detect agree with those obtained by Baudry \& Diamond
in sign and magnitude to better than 1~mG.

In addition to obtaining a positive field measurement for W3(OH), we
are able to place upper limits of several milligauss on $B_\parallel$
for two other sources: K3$-$50 and G10.624$-$0.385.  In K3$-$50 the
magntitude of the line-of-sight component deduced from the Zeeman
splitting of the main absorption feature is less than 2.9~mG.
Measurements of OH masers give three-dimensional magnetic field
strengths of $-2.6$ to $-7.5$~mG in the \threehalves\ lines
\citep{fishthesis} and $-5.3$ to $-9.1$~mG in the 6035~MHz $J = 5/2$
line \citep{baudry97}.

Our $3\sigma$ limit
of 1.2~mG for G10.624$-$0.385 improves by a factor of 2 on previous
observations in the OH \sevenhalves\ $F = \ttt$ line and is comparable
to results obtained from observations in SO absorption \citep{uchida}.
In G10.624$-$0.385, VLBA observations of the 1667~MHz $J = 3/2$ line
show one Zeeman pair with a magnetic field of $-6.0$~mG
\citep{fishthesis}.

Four possibilities, alone or in combination, can explain the
non-detection of a Zeeman pattern in the absorption in K3$-$50 and
G10.624$-$0.385.  (1) A reversal of the line-of-sight direction of the
magnetic field exists across each source.  (2) The magnetic field in
K3$-$50 and G10.624$-$0.385 is inclined at a large angle relative to the
line of sight.  (3) The average magnetic field strength of the
material sampled by \sevenhalves\ OH is smaller than that sampled by
$J = 5/2$ or $J = 3/2$ OH.  (4) Masers are on average denser than the
surrounding OH material, and thus the magnetic field strength is
greater at masing sites than for the OH as a whole.

The first explanation is possible for both these sources.  While five
\threehalves\ maser Zeeman pairs were found in VLBA imaging of K3$-$50,
they are all located on the periphery of the \hii\ region to the north
and east.  It is possible that the line-of-sight direction of the
magnetic field reverses on the south or west side of the source.  Even
if there is no reversal in K3$-$50, four of the five Zeeman pairs found
in the VLBA images of the 1665 and 1667 MHz maser emission imply full
(unprojected) magnetic field strengths less than the detectability
limits in Table \ref{zeeman-table} \citep{fishthesis}.

Only one ground-state OH maser has been found in VLBA imaging of
G10.624$-$0.385.  It would be a $12\sigma$ and $15\sigma$ detection
for the $F = \ttt$ and $\tff$ transitions, respectively, if the
average line-of-sight magnetic field component in the region of the $J
= 7/2$ absorption were the same as the full three-dimensional magnetic
field strength implied by Zeeman splitting of the 1667 MHz maser.  But
with only one Zeeman pair detected through synthesis imaging, it is
impossible to conclude whether or not there exists a line-of-sight
field reversal across the source.

The second explanation simply states that if the magnetic field is
inclined at a large angle to the line of sight, its projection along
the line of sight may be insufficient to produce a detectable Zeeman
splitting in Stokes V, for the case where the splitting is less than
the linewidth.  Assuming that the magnetic field strength of
6.0~mG for the single Zeeman pair detected with VLBI resolution in
G10.624$-$0.385 is a typical value for the source as a whole, the
required inclination angle of the magnetic field to produce a
line-of-sight component less than 1.2~mG is $> 78\degr$.  If the
magnetic field actually is inclined $78\degr$ to the line of sight,
the expected linear polarization fraction of the two
$\sigma$-components of the Zeeman pair is $\sin^2 78\degr /
(1 + \cos^2 78\degr) = 0.92$ \citep{gkk2}.  But no linear polarization
is detected in the maser \citep{fishthesis}, ruling out a high
inclination at the maser site, on the east side of the \hii\ region.
Linear polarization fractions consistent with inclination angles as
large as $76\degr$ are seen on the west side of the \hii\ region,
although most masers imply inclination angles much smaller than this
value.  Incomplete spatial coverage of OH masers across the source
render it difficult to estimate with certainty the inclination angle
averaged across the source, but it is almost certainly less than
$78\degr$.  Thus, while $B_\parallel < B = 6.0$~mG in G10.624$-$0.385,
the inclination angle is likely not large enough such that
$B_\parallel < 3\sigma = 1.2$~mG.  Still, an inclination angle smaller
than $78\degr$ in combination with another factor (such as a field
reversal across the source) may suffice to reduce the average
line-of-sight magnetic field strength below our detection threshold.

The third explanation is unlikely because a higher temperature is
required to populate the \sevenhalves\ states than for either the $J =
3/2$ or $J = 5/2$ states.  These higher temperatures likely require
that the distribution of OH in the $J = 7/2$ states be peaked closer
to the central heat source than for the $J = 3/2$ and $J = 5/2$
states, as is noted for masers in W3(OH) \citep{baudry98}.  Since the
density probably increases with decreasing radius, the density (and
therefore average field strength) of the material sampled by OH in the
$J = 7/2$ states should be higher than that of the lower states.
There is no evidence that the magnetic field strengths derived from
Zeeman splitting in $J = 7/2$ masers \citep{caswell04} are weaker than
those derived from masers in the $J = 5/2$
\citep{baudry97,caswellvaile} or $J = 3/2$ transitions
\citep[e.g.,][]{fram}.  Indeed, if W3(OH) is not atypical,
there is reason to believe that field strengths deduced from OH Zeeman
splitting actually increase as measured by higher excitation states in
the $^2\Pi_{3/2}$ ladder \citep{gusten,baudry98}.

The fourth explanation differs from the third in that it suggests that
masers may inherently occur at density enhancements in the surrounding
medium.  This is plausible given the conditions of formation of
interstellar OH masers.  Observed maser strengths require some
combination of OH enrichment and density enhancement
\citep{elitzurbook}.  OH masers are believed to form in the zone
between the ionization and shock fronts \citep{elitzurdejong}, where
instabilities lead to inhomogeneous density enhancement \citep[e.g.,
simulations of][]{garciasegura}.  The density conditions under which
OH masers form have implications which affect the physical
interpretation of their phenomenology.  If masers are formed
preferentially at density enhancements, magnetic field measurements
obtained at OH maser sites should be higher than the average magnetic
field strength in the surrounding region.  It would also be strong
evidence that observed proper motions of OH masers \citep[as
in][]{bloemhof} are due to discrete material motions, not shifting
coherence paths that could be unrepresentative of the motion of the
material.\footnote{The persistence of maser spot shapes also present a
strong case that proper motions are due to material motions
\citep{bloemhof96}.}

It is difficult to tell conclusively whether or not the material
sampled by $J = 7/2$ absorption is at a lower density than the
$^2\Pi_{3/2}$ masers in most of our sources.  In K3$-$50, the upper
limit on the magnetic field strength in the absorption falls within
the range obtained from masers.  In G10.624$-$0.385, the inclination
of the magnetic field to the line of sight alone is unlikely to
explain the discrepancy between our upper limit and maser field
strength values.  But it is quite possible that a field direction
reversal occurs across the source, which would reduce the effective
average line-of-sight strength detectable through absorption.  It is
also not possible to rule out that there exists a field direction
reversal \emph{along the line of sight}.  However, magnetic field
directions deduced from OH maser Zeeman splitting in massive
star-forming regions show an overwhelming tendency to fall into one of
two categories: (1) a constant line-of-sight direction throughout the
source, and (2) a single reversal across the source with the property
that a line can be drawn dividing the source into a region of positive
magnetic field and a region of negative magnetic field
\citep{fishthesis}.  A reversal of the field direction \emph{along the
line of sight} would produce projected regions of mixed magnetic field
direction, suggesting that they are not common in massive star-forming
regions.

As for W3(OH), the magnetic field measurement obtained from absorption
$B_\parallel = 3.0 \pm 0.3$~mG is less than those obtained from masers
projected atop the \hii\ region in the $^2\Pi_{3/2}, J = 3/2$
transitions \citep[average $\pm$ standard error of the mean: $B = 5.6
\pm 0.7$~mG, from][]{bloemhof}, $J = 5/2$ transitions \citep[$6.6 \pm
0.5$~mG, from][]{desmurs98}, as well as the $J = 7/2$ transitions
\citep[$9.7 \pm 0.4$~mG, from this work and][]{baudry98}.  The average
magnetic field measurements from the masers projected atop the \hii\
region exceed the absorption magnetic field measurement by $3.3\,
\sigma$, $5.8\, \sigma$, and $13.0\, \sigma$ for the $J = 3/2$, $5/2$,
and $7/2$ transitions, respectively.  Most masers in W3(OH) have no
detectable linear polarization \citep{garciabarreto}, suggesting that
the magnetic field in W3(OH) is oriented close to the line of sight
(i.e., $B_\parallel \approx B$) and therefore that the average
magnetic field sampled by $J = 7/2$ absorption is less than that
sampled by maser emission.



The average magnetic field strength obtained from the $J = 7/2$ masers
is greater than those obtained from $J = 3/2$ and $5/2$
masers.  As previously noted, the $J = 7/2$ masers in W3(OH) are much
more tightly distributed near the center of the source than in either
of the other masing $^2\Pi_{3/2}$ transitions.  Yet the magnetic field
strength measured from $J = 7/2$ absorption is significantly less than
the value obtained from $J = 7/2$ masers.  Taken together, these
results suggest that $J = 7/2$ masers occur only in the
highest-density portions of the OH cloud, even though OH likely exists
in the $J = 7/2$ excited state throughout the source.

Two possible scenarios could explain the prevalence of high-density
material necessary for excited-state OH maser activity.  There may be
a large region of higher density coincident with the distribution of
$J = 7/2$ and methanol masers \citep{moscadelli99} but with little
small-scale clumping.  Alternatively, small-scale ($\approx
10^{15}$~cm) clumping may occur throughout W3(OH) but with increased
prevalence in the region were $J = 7/2$ OH masers are observed.

Our data do not directly distinguish between these two scenarios.  Our
observations of Zeeman splitting in excited-state OH absorption allow
us to measure the magnetic field (and therefore density) averaged over
the entire source, but not the length scale on which density
fluctuations occur.  Nevertheless, other evidence suggests that
small-scale density variations are responsible for excited-state OH
maser activity.  The distribution of ground-state OH masers in
W3(OH) \citep{reid80} and other massive star-forming regions
\citep{fishthesis} show clustering on a scale of $10^{15}$~cm.  This
is unlikely to be caused by Kolmogorov turbulence, which is a
scale-free process.  Also, ammonia observations by \citet{rmb} show
that while the optical depth of NH$_3$ is fairly constant across
W3(OH), the beam-filling factor decreases from west to east.  This
suggests that the number of clumps is decreasing, not the density of
any one clump.

\begin{deluxetable}{lllll}
\tablewidth{0 pt}
\tablecaption{Absorption Parameters\label{absorption-table}}
\tablehead{ \colhead{Source} &
            \colhead{Transition} &
            \colhead{$v\lsr$} &
            \colhead{$T_A$} &
            \colhead{$\Delta v$} \\
            \colhead{} &
            \colhead{} &
            \colhead{(\kms)} &
            \colhead{(K)} &
            \colhead{(\kms)}
          }
\startdata
W3(OH)          & \ttt &      $-$47.22 & $-$0.177 &\phn3.75 \\
                &      &      $-$45.03 & $-$0.406 &\phn2.40 \\
                & \ttf &      $-$46.03 & $-$0.039 &\phn2.84 \\
                & \tft &      $-$44.71 & $-$0.013 &\phn3.40 \\
                & \tff &      $-$48.16 & $-$0.156 &\phn2.70 \\
                &      &      $-$44.99 & $-$0.620 &\phn3.25 \\
G10.624$-$0.385 & \ttt &   \phn$-$2.54 & $-$0.330 &\phn5.57 \\
                &      &\phmin\phn1.02 & $-$0.114 &\phn3.80 \\
                & \ttf &   \phn$-$2.36 & $-$0.030 &\phn6.64 \\
                & \tff &   \phn$-$1.86 & $-$0.536 &\phn6.13 \\
                &      &\phmin\phn0.99 & $-$0.047 &\phn1.89 \\
G28.199$-$0.048 & \ttt &   \phmin92.45 & $-$0.027 &\phn6.12 \\
                & \tff &   \phmin93.82 & $-$0.038 &\phn8.64 \\
W49             & \ttt &\phmin\phn4.34 & $-$0.152 &\phn4.11 \\
                &      &\phmin\phn6.22 & $-$0.219 &   21.69 \\
                & \ttf &\phn\phmin5.59 & $-$0.107 &   23.07 \\
                & \tff &\phmin\phn4.27 & $-$0.232 &\phn5.07 \\
                &      &\phmin\phn8.23 & $-$0.347 &   19.87 \\
K3$-$50         & \ttt &      $-$25.07 & $-$0.216 &\phn5.98 \\
                &      &      $-$20.10 & $-$0.040 &\phn2.90 \\
                & \ttf &      $-$25.50 & $-$0.018 &\phn7.11 \\
                & \tft &      $-$25.08 & $-$0.010 &\phn3.19 \\
                & \tff &      $-$25.02 & $-$0.276 &\phn6.34 \\
                &      &      $-$20.24 & $-$0.039 &\phn2.14 \\
ON 1            & \ttt &   \phmin10.88 & $-$0.014 &\phn7.13 \\
                & \tff &   \phmin10.74 & $-$0.016 &\phn7.34
\enddata
\end{deluxetable}

\begin{deluxetable}{lllll}
\tablecaption{Magnetic Field Measurements from Absorption and Maser
  Emission Lines\label{zeeman-table}}
\tablehead{
           \colhead{} &
	   \multicolumn{2}{c}{Absorption} &
           \colhead{} &
           \colhead{} \\
           \colhead{} &
           \colhead{\ttt} &
           \colhead{\tff} &
           \colhead{Masers\tnm{a}} &
           \colhead{} \\
           \colhead{Source} &
           \colhead{$B$ (mG)} &
           \colhead{$B$ (mG)} &
           \colhead{$B$ (mG)} &
           \colhead{References}
          }
\startdata
W3(OH)           & $3.0 \pm 0.3$ & $< 7.0$\tnm{b} & $5.6 \rarr 9.7$ & 1,2,3,4\\
G10.624$-$0.385  & $ < 1.5$      & $ < 1.2$ & $-6.0$ & 5 \\
G28.199$-$0.048  & $ < 60 $      & $ < 67 $ & $6.7$ & 6     \\
W49              & $ < 11 $      & $ < 7.8$ & $-4.3$ & 7     \\
K3$-$50          & $ < 2.9$      & $ < 3.1$ & $-3.7 \rarr -7.2$ & 5,8 \\
ON~1             & $ < 73 $      & $ < 97 $ & $-2.5 \rarr -6.0$ & 1,5,8
\enddata
\tnt{a}{Average magnetic field strengths deduced from Zeeman splitting
  in one or more $^2\Pi_{3/2}$ OH transitions are presented.  Note
  that magnetic field strengths determined from Zeeman splitting of OH
  masers are \emph{full} three-dimensional values, while those
  obtained from absorption are sensitive to only the
  \emph{line-of-sight projection} of the full magnetic field.  See \S
           \ref{zeeman} for more details.}
\tnt{b}{Contamination from strong masers and uncertainty in
  the strength and location of the absorption features precludes
  obtaining a lower estimate of the Zeeman splitting.}
\tablecomments{Upper limits are $3\sigma$ values.}
\tablerefs{
  (1) this work; (2) \citealt{bloemhof}; (3) \citealt{desmurs98}; (4)
  \citealt{baudry98}; (5) \citealt{fishthesis};
  (6) \citealt{caswellvaile}; (7) \citealt{caswell03};
  (8) \citealt{baudry97}}
\end{deluxetable}

\subsection{Line Ratios\label{line-ratios}}

In local thermodynamic equilibrium, the relative strengths of the $F =
\tft, \ttt, \tff,$ and $\ttf$ absorption lines are $1:27:35:1$.  The
ratio of the $F = \tff$ to $\ttt$ main lines ranges from 0.41
(G10.624$-$0.385, 1~\kms) or 0.97 to 1.62 in our sources, with 1.30
being the LTE value for optically thin lines.  The satellite lines,
when detected in absorption, are always enhanced relative to the
expected LTE value for the $F =
\tff$ line and almost always enhanced relative to the $F = \ttt$
transition, the single exception being the $F = \tft$ line in W3(OH).

\citet{matthews86} surmise that the excitation temperatures in the
hyperfine lines are not equal due to line overlaps in the 84~$\mu$m
lines connecting the \fivehalves\ and \sevenhalves\ states of OH.
\citet{viscuso85} note that collisions will preferentially excite the
\sevenhalves\ negative-parity state, but that the 84.42~$\mu$m line
that excites the positive-parity state may be in resonance with CO
emission at 84.41~$\mu$m.  Matthews et al.\ point out that far
infrared line overlaps will equalize populations between the $F = 4^+$
and $F = 3^+$ states, but the $F = 3^-$ state will depopulate relative
to the $F = 4^-$ state.  Hence, the $F = \tff$ and $\ttf$ absorption
lines will have a lower excitation temperature than the $F = \ttt$ and
$\tft$ lines.

For the two cases in which both satellite lines are detected, the $F =
\ttf$ line is stronger than the $F = \tft$ line.  Additionally, the
$\ttf$ line is detected in two sources in which the $\tft$ line is not
detected.  This is consistent with a lower excitation temperature for
the $\ttf$ state than for the $\tft$ state.  However, we do not find
evidence that the $F = \tff$ lines are enhanced relative to the $\ttt$
lines, as predicted by the Matthews et al.\ model.

\section{Conclusions}

We have detected \sevenhalves\ OH absorption toward six massive
star-forming regions.  Main-line absorption was detected in both the
$F = \ttt$ and $\tff$ lines toward all sources.  Additionally, we
detected at least one satellite line in absorption toward four of the
six sources.  We believe that this is the first detection of the
\sevenhalves\ satellite lines in interstellar sources.

Stokes V spectra of the main lines were produced for these six
sources.  In the case of W3(OH), Zeeman splitting of the $F = \ttt$
absorption results in a magnetic field measurement of $B_\parallel =
3.0 \pm 0.3$~mG, consistent in magnitude and sign with the $3.1 \pm
0.4$~mG obtained by \citet{gusten}.  W3(OH) is the only source in
which a positive detection of Zeeman splitting in the $J = 7/2$
absorption has been obtained.  The component of the magnetic field
along the line of sight is comparable to the full magnetic field
strength measured in $J = 3/2$ and $J = 5/2$ masers, suggesting that
these masers do not preferentially form in high density, low filling
factor regions
where the density significantly exceeds that of the surrounding,
non-masing OH.  However, the line-of-sight component of the magnetic field
determined from $J = 7/2$ absorption is much smaller than that
determined from $J = 7/2$ masers.

The grand question is whether the small line-of-sight magnetic field
strength measured in \sevenhalves\ absorption necessarily implies
density enhancement at maser sites.  If $J = 7/2$ absorption occurs
only where the $J = 7/2$ masers occur, then the full magnetic field
strength (as deduced from the $J = 7/2$ masers) in this region ranges
from 5.6 to 11.3~mG \citep{baudry98}, consistent with the range of
field strengths determined from $J = 3/2$ masers at this site at the
northern limb of the \hii\ region \citep{bloemhof}.  Restricting
consideration to the $F = \ttt$ transition, the transition in which a
positive result for Zeeman splitting is obtained in absorption, favors
the upper end of this range of field strengths, since the only
detected maser Zeeman pair implies a full magnetic field strength of
10.3~mG (Table \ref{maser-table}).  The magnetic field in this region
does not appear to be significantly inclined to the line of sight
\citep{garciabarreto}, suggesting that the line-of-sight magnetic
field strength is comparable to the full magnetic field strength.
Since the average magnetic field strength derived from masers in this
region is 9.1 mG, three times the field strength measured in
absorption, it might be concluded that masers are denser than the
surrounding, non-masing region by a factor of 9, under the reasonable
assumption that $B \propto n^{1/2}$.

However, it is likely that $J = 7/2$ absorption comes from other areas
in front of the \hii\ region as well.  \citet{bloemhof} identify three
other regions of \threehalves\ masers projected atop the \hii\ region
in W3(OH): a central clump with magnetic fields of 6.2 and 7.1~mG, a
southern clump with fields of 2.3 to 6.0~mG, and a western clump with
a magnetic field of 1.8~mG.  Observations of the $^2\Pi_{3/2}, J =
9/2$ transitions, with an excitation temperature of 511~K above
ground, show strong absorption at the northern clump as well as weaker
absorption at the central and southern clumps \citep{baudry95}.  In
the $^2\Pi_{1/2}, J = 3/2$ (270~K above ground) transitions,
significant absorption is seen at and between all four \threehalves\
maser clumps, while in the $^2\Pi_{1/2}, J = 5/2$ (415~K above ground)
transitions, absorption is seen mainly along a line running through
the northern, central, and southern clumps, with weaker absorption
from the western clump \citep{baudry93}.  There is no published map of
\sevenhalves\ (290~K above ground) absorption in W3(OH), but the
distribution of OH in other comparable excited states suggests that
strong \sevenhalves\ absorption would be seen over the majority of the
western half of the \hii\ region, including the sites of all four
\threehalves\ maser clumps.  This would imply that the density at
\threehalves\ maser sites is about 2 to 4 times that of the non-masing
OH.  On the other hand, 7820 MHz ($^2\Pi_{1/2}, J = 3/2, F = 2^+ \rarr
2^-$) and 8190 MHz ($^2\Pi_{1/2}, J = 5/2, F = 3^- \rarr 3^+$)
\emph{emission} is seen exclusively near the northern maser clump
\citep{baudry93}, and 6031 MHz (\fivehalves, $F = 2^- \rarr 2^+$) and
6035 MHz ($F = 3^- \rarr 3^+$) \emph{maser emission} is strongest near
the northern maser clump \citep{moran78,desmurs98}.  In total, this
suggests that while \sevenhalves\ absorption occurs over most of the
western half of W3(OH), the strongest contribution to the absorption
most likely comes from the northern clump, which has the highest
average magnetic field strength.  The unweighted average magnetic field
determined from OH masers in all transitions atop the \hii\ region is
6.9~mG.  Due to the distribution of maser spots, this effectively
gives highest weight to the northern clump and lowest weight to the
eastern clump, consistent with our arguments based on the distribution
of excited-state $^2\Pi_{1/2}$ and $^2\Pi_{3/2}$ OH absorption and
emission.  Thus, the density at maser sites is likely several
($\approx 5$) times that of the non-masing regions in W3(OH).  More
precise quantitative results would require an interferometric map of
\sevenhalves\ absorption.

We are able to place upper limits of several milligauss on
the line-of-sight component of the magnetic field in two other sources.
The $2.9$~mG limit on K3$-$50 is about 50\% of the full three-dimensional
magnetic field strengths obtained from Zeeman splitting of the $J =
3/2$ \citep{fishthesis} and $J = 5/2$ masers \citep{baudry97}.  The
$1.2$~mG limit on G10.624$-$0.385 is a factor of 5 smaller than
the $6.0$~mG obtained from a $J = 3/2$ maser Zeeman pair
\citep{fishthesis}.  It is not clear if this occurs because the masers
are in higher density (and hence higher magnetic field) clumps or if
other effects, such as field reversals or large angles of the magnetic
field to the line of sight, are present in portions of the source.

We have detected $F = \ttf$ and/or $\tft$ satellite line absorption
in four sources.  Of these two, absorption in the $\ttf$ line is
always stronger, as predicted by \citet{matthews86}.  Satellite line
absorption appears enhanced relative to the main lines, from
that which would be expected in local thermodynamic equilibrium.

Maser emission was also observed toward two sources.  In W3(OH) we
find seven pairs of LCP and RCP masers implying magnetic field
strengths from $6.9$ to $11.3$~mG.  We detect line components
at all velocities where strong maser features were
previously observed, although we do not have the spatial resolution to
separate the multiple lines detected by \citet{baudry98} using the
VLBA.  In addition, we find a pair of previously undetected masers
centered at $-44.85$~\kms.  In ON~1, we find the maser at $14$~\kms\
previously detected by \citet{baudry02}.  We do not see the strong
maser at $-0.13$~\kms\ that they did, but we find a new weak maser
centered at $0.31$~\kms.  Maser strength variability, previously noted
in the \sevenhalves\ masers of W3(OH) \citep[e.g.,][]{baudry98},
appears to be operating in ON~1 as well.

\acknowledgments
Support for this work was provided by the NSF through
award GSSP04-0001 from the NRAO.  We thank T.~Minter for help in
setting up the observations, J.~Braatz and G.~Langston for assistance
in data reduction, and M.~Modjaz for helpful comments in the
interpretation of the data.  We also thank an anonymous referee for
helpful comments.

\clearpage

\clearpage

\end{document}